\documentclass[twocolumn,showpacs,preprintnumbers,amsmath,amssymb]{revtex4}
\usepackage{latexsym}% 
\usepackage{graphicx}% Include figure files
\usepackage{dcolumn}% Align table columns on decimal point
\usepackage{bm}% bold math
\begin{document}

\voffset= 1.0 truecm 

%\preprint{APS/123-QED}
%
\newcommand{\bea}{\begin{eqnarray}}
\newcommand{\eea}{\end{eqnarray}}
\newcommand{\be}{\begin{equation}}
\newcommand{\ee}{\end{equation}}
%
%%%%%%%%%%%%%%%%%%  BOLDFACE GREEK LETTERS   %%%%%%%%%%%%%%%%
\newcommand{\xbf}[1]{\mbox{\boldmath $ #1 $}}
%%%%%%%%%%%%%%%%%%%%%%%%%%%%%%%%%%%%%%%%%%%%%%%%%%%%%%%%%%%%%

\title{Electromagnetic multipole moments of baryons}
\thanks{Published in Few-Body Syst. {\bf 59}, 145 (2018).}
\author{Alfons J. Buchmann}
\affiliation{Institute for Theoretical Physics,
University of T\"ubingen, D-72076 T\"ubingen, Germany}
\email{alfons.buchmann@uni-tuebingen.de}

\pacs{13.40.Em, 13.40.Gp, 14.20.-c, 11.30.Ly}

\begin{abstract}
We calculate the charge quadrupole and magnetic octupole 
moments of baryons using a group theoretical approach based 
on broken SU(6) spin-flavor symmetry.
The latter is an approximate symmetry of the QCD Lagrangian 
which becomes exact
in the large color $N_c$ limit. Spin-flavor symmetry breaking is induced 
by one-, two-, and three-quark terms in the electromagnetic current operator.  Two- and three-quark currents provide the leading contributions for higher multipole moments,
despite being of higher order in an $1/N_c$ expansion.
Our formalism leads to relations between $N \to N^*$ transition multipole moments and nucleon ground state properties. We compare our results to experimental quadrupole and octupole transition moments extracted from measured helicity amplitudes.
\end{abstract}

\maketitle

\section{Introduction}
\label{intro}

Electromagnetic multipole moments of baryons are interesting observables. 
They are directly connected with the spatial charge and current distributions 
in baryons, and thus contain information about their size, shape, and 
internal structure. In particular, charge quadrupole and magnetic octupole moments provide important information on the geometric shape of baryons, which is not available from the corresponding leading multipole moments.

However, higher electromagnetic multipole moments, such 
as charge qua\-dru\-pole (C2) and magnetic octupole (M3) moments of spin
 $\ge 3/2$ baryons are very difficult to measure.
Presently, we have no direct experimental information on these moments, 
but it is planned to measure the quadrupole moment of 
the $\Omega^-$ baryon at FAIR  
in Darmstadt\cite{Poc17}. 
This is contrasted by several theoretical works on baryon quadrupole 
moments~\cite{But94,Leb95,Oh95,Hen01,Hes02,Dah13,Kri91,Ram09,Gia90} 
and relatively few on magnetic octupole moments~\cite{Ram09,Gia90,Buc08,Ali09}. 

\begin{figure*}
\begin{center}
  \includegraphics[width=0.55\textwidth]{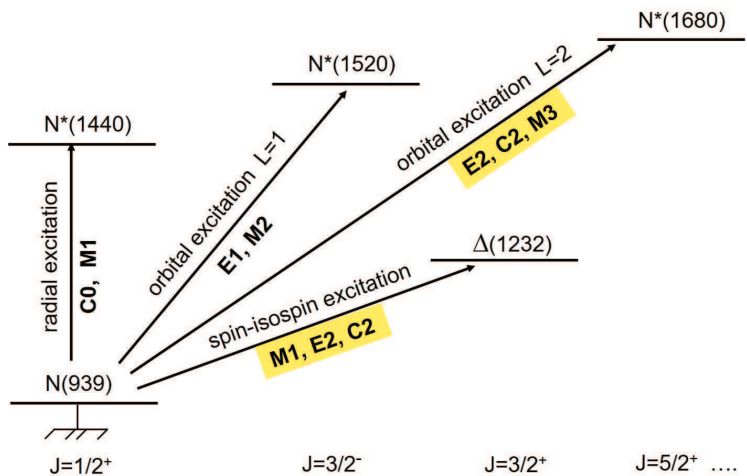}
% figure caption is below the figure
\caption{Low-lying nucleon resonances and transition multipole moments M, E, C.}
\label{fig:1}       % Give a unique label
\end{center}
\end{figure*}

On the other hand, transition multipole moments between 
the ground state and excited
states of the nucleon as shown in Fig.~\ref{fig:1} 
are experimentally accessible. 
High precision electron and 
photon scattering experiments, exciting the lowest lying nucleon resonance 
$\Delta^+(1232)$ have provided evidence for a nonzero $p\to \Delta^+$ 
transition quadrupole moment $Q_{p \to \Delta^+}$  
and hence for a nonspherical charge distribution in baryons. 
The experimental results~\cite{Tia03,Bla01} are in agreement with the quark model prediction~\cite{BHF97}
\be
\label{qmnr}
Q_{p \to \Delta^+}= \frac{1}{\sqrt{2}}\, r_n^2, 
\ee
where $r_n^2$ is the neutron charge radius~\cite{Kop97}. It has been suggested 
that a transition quadrupole moment of the sign as in Eq.(\ref{qmnr}) arises 
because the proton has a prolate 
and the $\Delta^+$ an oblate charge distribution and that the neutron charge radius
is a measure of the intrinsic quadrupole moment of the nucleon~\cite{Hen01}.
For reviews see Ref.~\cite{Pas07,Ber07}. 

Furthermore, it was proposed~\cite{Buc04} that Eq.(\ref{qmnr}) 
is the zero momentum transfer limit of a more general relation between 
the $p\to \Delta^+$ charge quadrupole 
transition form factor $G_{C2}^{p \to \Delta^+}(Q^2)$ and the elastic neutron charge 
form factor $G_{C}^n(Q^2)$
\be
\label{C2C0rel}
G_{C2}^{p \to \Delta^+}(Q^2)=-\frac{3 \sqrt{2}}{Q^2}\, G_{C}^n(Q^2).
\ee
Eq.(\ref{C2C0rel}) agrees with experiment for a wide range of momentum transfers.
In addition, it has the correct low $Q^2$ behavior of a charge quadrupole form factor, 
and the correct high $Q^2$ asymptotic behavior predicted by perturbative QCD~\cite{Idi04,Tia07,tia16}.

The purpose of the present contribution is to further explore relations between transition multipole moments and nucleon ground state properties. 
We will focus our attention on the Coulomb quadrupole (C2) and magnetic octupole (M3) transition form factors as shown in Fig.~\ref{fig:1}. 
The reason for this is that
these are next-to-leading moments of the elastic charge monopole (C0) and magnetic
dipole (M1) nucleon form factors.  While the nucleon ground state does not have
spectroscopic quadrupole and octupole moments, it does have corresponding intrinsic
moments. 
We will extract the sign and size of these intrinsic ground state moments from the measurable transition moments and discuss their implications for the shape of the nucleon.

% For one-column wide figures use
%\begin{figure}
% Use the relevant command to insert your figure file.
% For example, with the graphicx package use
% \includegraphics{spectrum.eps}
% figure caption is below the figure
%\caption{Low-lying excited states of the nucleon}
%\label{fig:1}       % Give a unique label
%\end{figure}
%
% For two-column wide figures use

%%%%%%%%%%%%%%%%%%%%%%%%%%%%%%%%%%%%%%%%%%%%%%%%%%%%%%%%%%%%%%%%%%%%%%%%%
%             FIG. 1: Low- ying excited nucleon states 
%%%%%%%%%%%%%%%%%%%%%%%%%%%%%%%%%%%%%%%%%%%%%%%%%%%%%%%%%%%%%%%%%%%%%%%%%

\section{Electromagnetic excitation of nucleon resonances}
\subsection{Elastic and inelastic electron-nucleon scattering}
\begin{figure*}
\begin{center}
\includegraphics[width=0.85\textwidth]{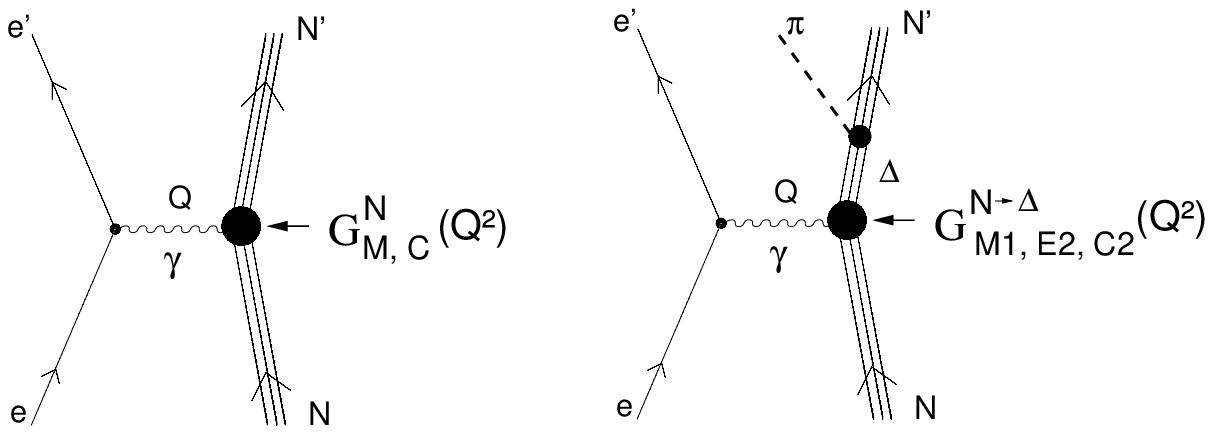}
\caption{\label{fig:scattering} 
Left: Elastic electron-nucleon 
scattering $e\, N \to e' \, N'$ 
involving the exchange of a single virtual photon $\gamma$ of 
four-momentum $Q$, with $Q^2=-(\omega^2 -{\bf q}^2)$. 
Here, $\omega$ is the energy transfer and ${\bf q}$ the three-momentum transfer 
of the virtual photon.
The nucleon structure information is contained in 
the charge monopole form factor $G^{N}_{C}(Q^2)$ and
magnetic dipole form factor
$G^{N}_{M}(Q^2)$.
Right: Inelastic electron-nucleon scattering $e \, N \to e' \, \Delta$. 
The excitation of the $\Delta$ resonance is described 
by three electromagnetic transition form factors 
$G^{N \to \Delta}_{M1}(Q^2)$, $G^{N \to \Delta}_{E2}(Q^2)$, 
and $G^{N \to \Delta}_{C2}(Q^2)$. }
\end{center}
\end{figure*}

On the way towards reaching a better understanding of the nucleon spectrum,
meson electroproduction experiments depicted 
in Fig.~\ref{fig:scattering}(right) have been particularly fruitful.
For recent reviews see~\cite{Tia11,Azn11}.
From the theory side, the electromagnetic structure of the nucleon is described in terms of elastic and inelastic multipole form factors as indicated 
in Fig.~\ref{fig:scattering}.
The elastic form factors can to a certain extent be interpreted as Fourier transforms 
of the charge $\rho({\bf r})$ and spatial current distributions ${\bf J}({\bf r})$ 
inside the nucleon.
Thus, the elastic form factors are directly related to the spatial 
structure of the nucleon ground state, which in turn determines the transition form
factors to excited nucleon states. 

\begin{figure*}
\begin{center}
  \includegraphics[width=0.85\textwidth]{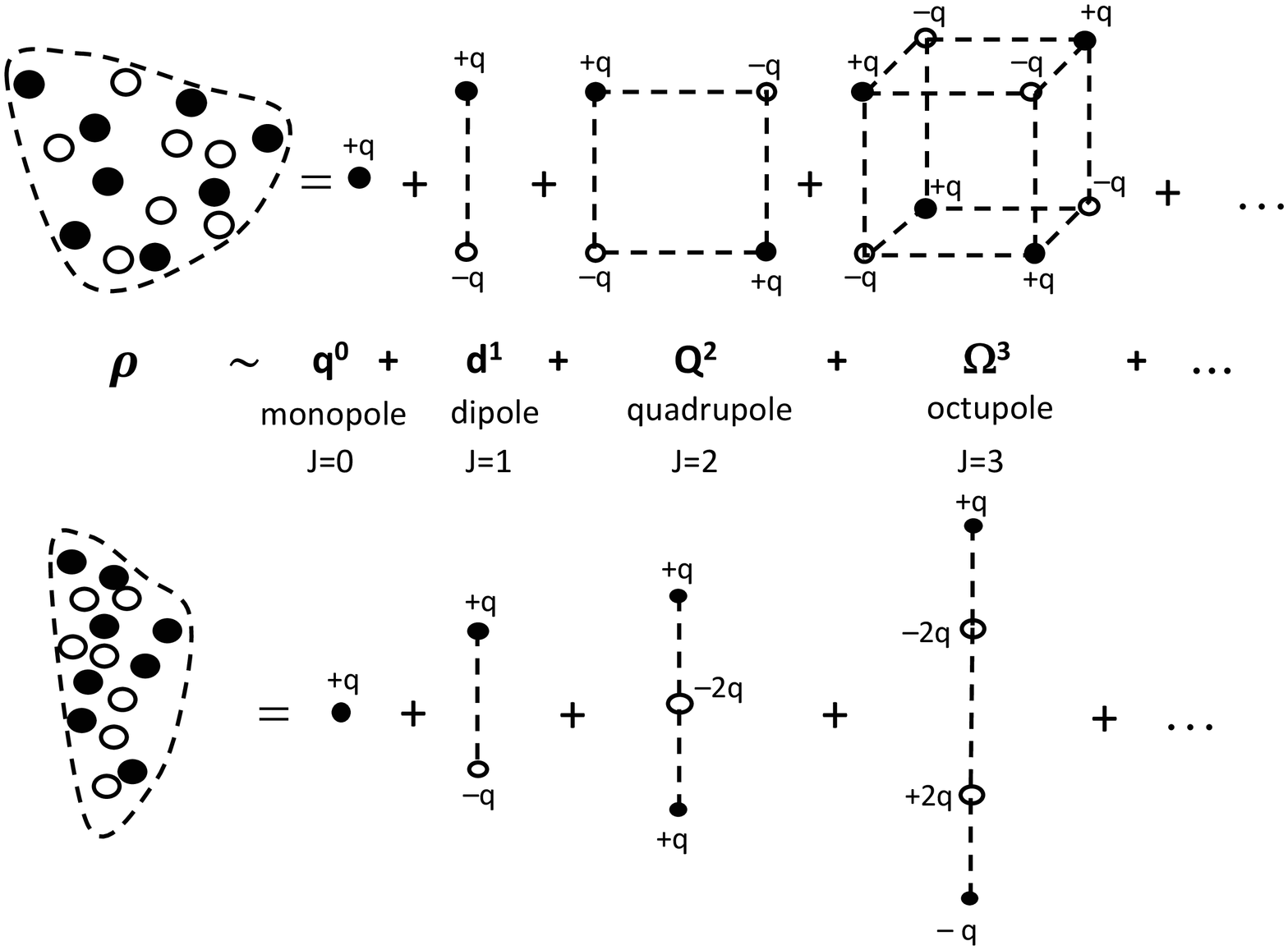}
% figure caption is below the figure
\caption{Qualitative illustration of the multipole expansion of classical charge densities $\rho$ into a series of Coulomb $2^J$-poles of multipolarity $J$.
The different geometric shapes of $\rho$ are reflected 
by the sign and size of the higher multipole moments. 
Upper panel: oblate (pancake-shaped) $\rho$ with intrinsic quadrupole moment $Q_0 <0$. 
Lower panel: prolate (cigar-shaped) $\rho$ with $Q_0>0$. 
Due to parity and time reversal invariance of the electromagnetic interaction 
only Coulomb multipole moments with even multipolarity are allowed quantum mechanically.}
\label{fig:multipoles}       % Give a unique label
\end{center}
\end{figure*}

Quite generally, the geometric properties of the ground state of a physical 
system, in particular its size and shape, have a direct bearing on the eigenfrequencies
 and eigenmodes of its excitation spectrum. Conversely, 
knowledge of the eigenfrequencies and excitation modes of a system enables 
us to draw certain conclusions concerning its size and shape. This also applies to 
the nucleon and suggests that e.g. the $N \to \Delta$ charge quadrupole (C2) 
transition form factor provides details about the nucleon ground state structure 
such as the quadrupole part of the charge density qualitatively illustrated
in Fig.~\ref{fig:multipoles}.

When trying to make inferences about the structure of a physical system based
on the excitation spectrum and transition multipoles to various excited states, 
the symmetries respected by the system provide valuable guidance. 
The regularities seen in the excitation spectrum and other observables 
of a quantum mechanical system are usually due to an underlying symmetry 
and thus call for a group-theoretical 
treatment. An early example is the explanation of the orbital angular 
momentum $l$ degeneracy and the $1/n^2$ law in the spectrum of atomic hydrogen 
by Pauli and Fock~\cite{Pau26}. Both properties were shown to follow from an underlying 
SO(4)$\sim$SU(2)$_V\times$SU(2)$_A$ symmetry that is 
isomorphic to the direct product of two SU(2) groups connected with two conserved quantities, orbital angular momentum (A) and the Lenz-Runge vector (V)~\cite{Gre94}.

In the case of baryons, SU(2) isospin symmetry, 
as well as the higher flavor SU(3)$_F$ and spin-flavor SU(6)$_{SF}$ symmetries 
and their breaking provide useful guidelines not only for the classification of states 
but also for extracting information on baryon structure from electromagnetic
multipoles. We will discuss the symmetry properties of electromagnetic multipoles 
in some detail in Sect.~\ref{sec:multop} and Sect.~\ref{sec:multisym}. 
%%%%%%%%%%%%%%%%%%%%%%%%%%%%%%%%%%%%%%%%%%%%%%%%%%%%%%%%%%%%%%%%%%%%%%%%%
%             FIG. 2:Elastic and Inelastic electron nucleon scattering
%%%%%%%%%%%%%%%%%%%%%%%%%%%%%%%%%%%%%%%%%%%%%%%%%%%%%%%%%%%%%%%%%%%%%%%%%
%%%%%%%%%%%%%%%%%%%%%%%%%%%%%%%%%%%%%%%%%%%%%%%%%%%%%%%%%%%%%%%%%%%%%
\subsection{Multipole operators and form factors}
\label{sec:multop}
Baryons are quantum mechanical systems with definite angular momentum and parity. 
It is therefore advantageous to describe their electromagnetic interaction in terms of electromagnetic multipole operators which transfer definite angular momentum $J$ and parity.
Angular momentum and parity selection rules then greatly facilitate the evaluation 
of matrix elements. Usually, only a few multipoles suffice to obtain a satisfactory description of the charge $\rho$ and current ${\bf J}$ distributions of the system.

A multipole expansion of the baryon charge density $\rho({\bf q})$ 
into Coulomb multipole operators $T_M^{[C] J}(q)$ is then given as~\cite{def66}
\be
\label{cha_mult_exp}
\rho({\bf q})= 4 \pi \sum_{J M} i^J Y^J_M({\bf{\hat{q}}}) \, T_M^{[C] J}(q).
\ee
Here, ${\bf q}$ is the three-momentum transfer of the virtual photon and 
$Y^J_M({\bf{\hat{q}}})$ is a spherical harmonic of multipolarity $J$ and projection $M$.
The Coulomb multipole operator $T_M^{[C] J}(q)$ is a spherical tensor of rank $J$ and parity $(-1)^J$
that is calculated from the charge density as follows
\be
\label{cmo}
T_M^{[C] J}(q)= \int j_J(qr) Y^J_M({\bf{\hat{r}}}) \rho({\bf r}) 
\, d^3{\bf r},
\ee
where $j_J(qr)$ is a spherical Bessel function of order $J$.

Analogously, the transverse current density is expanded 
into transverse electric $ T_{\lambda}^{[E] J}(q)$ and 
magnetic $T_{\lambda}^{[M] J}(q)$ multipole operators, which are spherical tensors 
of rank $J$ with parity $(-1)^J$ and $(-1)^{J+1}$ respectively as~\cite{def66} 
\be
\label{curmo}
{\bf J}_{\lambda}(q)= -\sqrt{2\pi} \sum_{J\ge 1} (-i)^J {\hat{J}}
\left[\lambda \, T_{\lambda}^{[M] J}(q)+ T_{\lambda}^{[E] J}(q) \right],
\ee
where $\lambda$ can take on the values $\lambda=\pm 1$ and ${\hat{J}}=\sqrt{2 J+1}$.
The transverse magnetic and electric multipole operators
are defined in terms of the spatial current density as 
\bea
\label{mmo}
T_{\lambda}^{[M] J}(q)\! \!&=& \!\!\int \! j_J(qr) {\bf Y}^{(J1)J}_{\lambda}({\bf{\hat{r}}}) \cdot {\bf J}({\bf r}) \, d^3{\bf r} \nonumber \\
T_{\lambda}^{[E] J}(q) \! \!&=& \! \!\frac{1}{q} 
\int \! {\xbf \nabla} \times (j_J(qr) {\bf Y}^{(J1)J}_{\lambda}({\bf{\hat{r}}})) \cdot {\bf J}({\bf r}) \, d^3{\bf r},
\eea
where ${\bf Y}^{(J1)J}_{\lambda}({\bf{\hat{r}}})$ are vector spherical harmonics.
%%%%%%%%%%%%%%%%%%%%%%%%%%%%%%%%%%%%%%%%%%%%%%%%%%%%%%%%%%%%%%%%%%%%%%%%%
%             FIG. 3: Multipole expansion of classical charge density
%%%%%%%%%%%%%%%%%%%%%%%%%%%%%%%%%%%%%%%%%%%%%%%%%%%%%%%%%%%%%%%%%%%%%%%%%
%%%%%%%%%%%%%%%%%%%%%%%%%%%%%%%%%%%%%%%%%%%%%%%%%%%%%%%%%%%%%%%%%%%%%%%%%%%
With angular momentum $J_i$ in the inital state and $J_f$ in the final state, 
angular momentum conservation restricts the number of multipole form factors 
of multipolarity $J$ as 
\be 
\vert J_i -J_f \vert \leq  J \leq J_i + J_f. 
\ee
Furthermore, parity and time reversal invariance of the electromagnetic interaction implies that in elastic scattering, there can be only even charge multipoles and odd magnetic multipoles but no transverse electric multipoles.
Specifically, for the positive parity nucleon ground state $N(939)$,
the $\Delta(1232)$, and the $N^*(1680)$ resonance the allowed 
elastic and transition multipoles are listed in Table~\ref{CME}.
%%%%%%%%%%%%%%%%%%%%%%%%%%%%%%%%%%%%%%%%%%%%%%%%%%%%%%%%%%%%%%%%%%%%%%%%%%%%%%%%%%%%%
\begin{table}[htb]
\begin{center}
\caption[Multipoles]{Coulomb (C), magnetic (M) and electric (E) multipoles
of multipolarity $J$ in elastic and inelastic electron-nucleon scattering for selected positive parity states. } 
\label{CME}
\begin{tabular}{| l | c |  c |} 
\hline
state & elastic  & transition   \\
\hline 
$N(939)$       &  C0, M1	     &   ---        \\
               &                 &              \\
$\Delta(1232)$ &  C0, C2         & C2           \\
               &  M1, M3     	 & M1, E2       \\
               &                 &              \\
$N^{*}(1680)$  & C0, C2, C4      &  C2          \\
               & M1, M3, M5  	 &  E2, M3      \\ \hline
\end{tabular} 
\end{center}
\end{table}
%===========================================================================

In general, the multipole operators depend on the photon three-momentum transfer 
$\vert {\bf q}\vert=q $, multipolarity $J$ and projection $M$. 
Their matrix elements give rise to corresponding transition multipole form factors~\cite{def66}
\bea
\label{mff}
G^{N\to N^*}_{C0}\!\!(q^2) & \! \! \! = \! \! \! & \sqrt{4 \pi}\,  
\langle N^* \vert T_0^{[C0]}(q) \vert N  \rangle,  \nonumber \\
G^{N\to N^*}_{C2}\!\!(q^2) &\! \!\! = \!\! \!& \frac{12 \sqrt{5 \pi}}{q^2}\,  
\langle N^* \vert T_0^{[C2]}(q) \vert N  \rangle,  \nonumber \\
\tilde{G}^{N\to N^*}_{M1} \!\!(q^2)& \!\! \!=\! \!\! & 
\frac{i \sqrt{6 \pi}}{q} (2M_N)\, 
\langle N^*\vert T_0^{[M1]}(q) \vert N \rangle, \nonumber \\
\tilde{G}^{N\to N^*}_{M3}\!\!(q^2) &\! \!\! =\! \!\! & 
-\frac{ i 15 \sqrt{21 \pi}}{q^3} (2M_N)
\langle N^*\vert T_0^{[M3]}(q) \vert N \rangle,
\eea
where $N^*=N$ (elastic) and $N^*=\Delta(1232)$ or $N^*=N^*(1680)$ (inelastic).
By convention, elastic and inelastic multipole form factors are evaluated for the $M=0$ projection of the multipole operator and the highest allowed total angular momentum projection of the baryon states involved. 

In the present paper, we focus on the static {\it multipole moments}, 
which are the $q \to 0$ limit of the multipole form factors in Eq.(\ref{mff}).
In this limit our multipole form factors are normalized to the usual spherical 
multipole moments known from classical electrodynamics~\cite{Jac75} 
and are straightforward generalizations of the Sachs form factor normalization used in elastic scattering.
In the $q \to 0$ limit we obtain from Eq.(\ref{mff}) and the definitions 
in Eq.(\ref{cmo}) and Eq.(\ref{mmo}) the total charge ${\rm Q}_N$ and magnetic moment 
$\tilde{{\mu}}_N$ in the elastic 
scattering case ($N^*=N$), in addition to the charge monopole and quadrupole as well as the magnetic dipole and octupole transition moments in inelastic scattering
\bea
\label{C2andM3}
{\rm{Q}}_{N \to N^*} \! \! & \! \!\!\! =\! \! \!& G_{C0}^{N \to N^*}(0)\!=\!\langle N^* \vert \!\! \int \! \rho({\bf r})\, d^3{\bf r} \, \vert N \rangle \nonumber \\
Q_{N \to N^*} \! \! & \! \! \!\!=\!\! \! & G_{C2}^{N \to N^*}(0)\!=\!\langle N^* \vert \!\! \int \! \rho({\bf r})\, (3 z^2 - r^2)\, d^3{\bf r} \, \vert N \rangle \nonumber \\
\tilde{\mu}_{N \to N^*}\! \!  & \! \!\!\!=\!\! \! &  \tilde{G}_{M1}^{N \to N^*}(0) \!= \!\frac{1}{2} 
(2M_N)\langle N^* \vert \! \!  \int \! ({\bf r} \times 
{\bf J}({\bf r}))_z \, d^3{\bf r} \, \vert N \rangle  \nonumber \\
\tilde{\Omega}_{N \to N^*} \! \! 
& \! \!\!\!=\! \!\!&  \tilde{G}_{M3}^{N \to N^*}(0) \nonumber \\
& \! \! =  \! \!& \frac{3}{8} (2M_N)
\langle N^* \vert \! \!  \int \! ({\bf r} \times 
{\bf J}({\bf r}))_z  (3 z^2 - r^2)  d^3{\bf r}  \vert N \rangle.  
\eea 
Defining $\mu_{N \to N^*}:=\tilde{\mu}_{N \to N^*}/(2M_N)$ and
$\Omega_{N \to N^*}:=\tilde{\Omega}_{N \to N^*}/(2M_N)$, which are then expressed in units of nuclear magnetons.

Other definitions of inelastic form factors with different normalizations 
have been written by several authors~\cite{jon73,dek76,war90,car86}. 
The advantage of the generalized transition Sachs form factors 
in Eq.(\ref{mff}) is that they are based on the same definition 
of the multipole operators that are used for the elastic Sachs form factors. This facilitates the comparison between elastic and inelastic nucleon form factors.

In the next section, we study the implications of broken SU(6) spin-flavor symmetry
for electromagnetic multipole moments and the reasons for the existence of   
relations between elastic and inelastic electromagnetic form factors such as 
Eq.(\ref{qmnr}) and its generalization to finite 
momentum transfers in Eq.(\ref{C2C0rel}).

\section{Multipoles from broken SU(6) spin-flavor symmetry}
\label{sec:multisym}
\subsection{SU(6) spin-flavor symmetry and its breaking }
\label{sec:spin-flavor}
It is well known that spin-flavor SU(6) symmetry unites the
spin 1/2 flavor octet baryons ($2 \times 8$ states) 
and the spin 3/2 flavor decuplet baryons ($4 \times 10$ states) 
into a common ${\bf 56}$ dimensional mass degenerate 
supermultiplet. If SU(6) spin-flavor symmetry were exact, octet and 
decuplet masses would be degenerate, baryon magnetic moments would be 
proportional to $\mu_p$, and baryon quadrupole moments as well as 
the charge radii of neutral baryons would be zero.

In nature, spin-flavor symmetry is broken. Due to SU(6) symmetry breaking 
the ${\bf 56}$ dimensional baryon supermultiplet
decomposes into irreducible representations of the SU(3) flavor 
and SU(2) spin subgroups of SU(6) as follows
\be
{\bf 56} = ({\bf 8}, {\bf 2}) + ({\bf 10}, {\bf 4 }), 
\ee
where the first and second entry in the parentheses indicate the dimension of the 
flavor and spin representations. The latter is given by $2J+1$.

For a general SU(N) group the symmetry breaking operators are constructed from the $N^2-1$ generators of the group SU(N). In particular, 
the 35 generators  of SU(6) are composed 
of 3 spin generators, 8 flavor generators, and 24 spin-flavor generators 
\be
\label{SU(6)_generators}	
\xbf{\sigma}_i,  \ \ \xbf{\lambda}_{\alpha}, \ \ 
\xbf{\sigma}_i\,\xbf{\lambda}_{\alpha}
\ee
with spin index $i=1,2,3$ and flavor index $\alpha=1,\cdots,8$.
These generators transform according 
to the adjoint or regular ${\bf 35}$ representation of SU(6).
Each generator stands for a different direction in a 35 dimensional 
vector space and breaks SU(6) symmetry in a specific way. 
%%%%%%%%%%%%%%%%%%%%%%%%%%%%%%%%%%%%%%%%%%%%%%%%%%%%%%%%%%%%%%%%%%%%%%%%%%%%%%%%%%

The transformation properties of the allowed spin-flavor symmetry breaking operators 
are then derived from group theory as follows. Using Littlewood's theorem, 
one decomposes the
product representation $\overline{{\bf 56}} \times {\bf 56}$ arising in 
matrix elements of an operator $\mathcal{O}^R$ between baryon states 
\be
\mathcal{M} = \langle  {\overline {\bf 56}} \vert \mathcal{O}^R \vert {\bf 56} \rangle
\ee
into irreducible SU(6) representations. An allowed operator $\mathcal{O}^R$ 
must transform according to one of the irreducible representations (irreps) $R$ 
found in the product~\cite{Gur64}
\be
\label{baryon56}
\overline{\bf 56} \times {\bf 56} 
=  {\bf 1} + {\bf 35} + {\bf 405} + {\bf 2695}.
\ee 
Operators transforming according to other SU(6) representations $R$
not contained in this product will lead to vanishing matrix elements 
when evaluated between states belonging to the ${\bf 56}$. 

The SU(6) dimension $R$ of an operator determines the operator type.
In particular, the {\bf 1} dimensional representation is associated 
with a zero-body operator (constant), whereas the ${\bf 35}$, ${\bf 405}$, and 
${\bf 2695}$ dimensional representations, 
are respectively connected with one-, two-, 
and three-quark operators~\cite{Leb95}. The corresponding spin-flavor operators are also 
refered to as SU(6) symmetric, and as first, second, and third order SU(6) symmetry 
breaking operators.

One-quark operators transforming according to the ${\bf 35}$ dimensional adjoint
representation of SU(6) cannot generate nonzero neutral baryon charge radii
and nonzero qua\-dru\-pole moments. In the case of quadrupole moments, this is seen 
after decomposing the ${\bf 35}$ dimensional representation into a sum of direct 
products of irreps of the SU(3)$_F$ and SU(2)$_J$ subgroups of SU(6) as
\be
\label{35decomp}
{\bf 35} = ({\bf 8},{\bf 1}) + ({\bf 8},{\bf 3}) + ({\bf 1},{\bf 3}).
\ee
 Clearly, the ${\bf 35}$ irrep does not contain 
a 5 dimensional representation in spin space necessary for a spin tensor of 
rank $J=2$ tensor such as the quadrupole moment operator. Therefore,
first order SU(6) symmetry breaking one-quark operators cannot produce nonvanishing quadrupole moments.

For later reference, we reproduce here the SU(3)$_F\times$SU(2)$_J$ decompositions 
for the second and third order SU(6) symmetry breaking operators~\cite{beg64a,Gou67} 
\bea
\label{405decomp}
{\bf 405} & = & ({\bf 1},{\bf 1}) + ({\bf 1},{\bf 5}) \nonumber \\ 
&+& ({\bf 8},{\bf 1}) + 2 \, ({\bf 8},{\bf 3}) + ({\bf 8},{\bf 5}) \\
&+& ({\bf 10},{\bf 3}) +  ({\overline {\bf 10}},{\bf 3})  + ({\bf 27},{\bf 1}) + 
({\bf 27},{\bf 3}) + ({\bf 27},{\bf 5}) \nonumber.  
\eea
\bea
\label{2695decomp}
{\bf 2695} & = & ({\bf 1},{\bf 7}) + ({\bf 1},{\bf 3})  \nonumber \\ 
&+ &  ({\bf 8},{\bf 7}) + 2 ({\bf 8},{\bf 5}) +  2({\bf 8},{\bf 3}) +
({\bf 8},{\bf 1}) \nonumber \\
&+ & ({\bf 10},{\bf 5})+( {\overline {\bf 10}},{\bf 5})+({\bf 10},{\bf 3})
\nonumber \\ 
& + & ({\overline {\bf 10}},{\bf 3})+( {\bf 10},{\bf 1})+
({\overline {\bf 10}},{\bf 1}) \nonumber \\
&+ & ({\bf 27},{\bf 7}) +  2({\bf 27},{\bf 5}) + 3({\bf 27},{\bf 3}) + 
({\bf 27},{\bf 1}) \nonumber \\
&+ &  ({\bf 35},{\bf 5}) +  ( {\overline {\bf 35}},{\bf 5}) + 
({\bf 35},{\bf 3}) + ( {\overline {\bf 35}},{\bf 3}) \nonumber \\
& +&  ({\bf 64},{\bf 7}) +  ({\bf 64},{\bf 5}) + ({\bf 64},{\bf 3}) + 
({\bf 64},{\bf 1}). 
\eea

Why is all this relevant for calculating electromagnetic multipoles? 
There are at least two reasons for this.
First, spin-flavor decompositions of SU(6) representations 
as in Eq.(\ref{405decomp}) allow the identification of a given multipole 
with a specific SU(3)$_F\times$SU(2)$_J$ product representation using the following rules.

{\bf Rule 1:} In lowest order of SU(3)$_F$ symmetry breaking, electro\-magnetic 
multipoles must transform according to the ${\bf 8}$ dimensional regular 
(or adjoint) representation of SU(3)$_F$ pertaining to the 8 generators of
SU(3)$_F$ because electromagnetic multipoles contain the electric charge $\rm{Q}$, 
which according to the Gell-Mann-Nishijima relation is built from the SU(3) 
generators $T_3=\lambda_3/2$ (isospin) and $Y=\lambda_8/\sqrt{3}$
(hypercharge)
\be
\label{GMNR}
{\rm{Q}} = T_3 + \frac{1}{2} Y= 
\frac{1}{2}\left (\lambda_3 + \frac{1}{\sqrt{3}}\lambda_8 \right ). 
\ee
Here, $T_3$ is the third component of isospin and $Y$ is the hypercharge.
More general flavor operators containing second and third 
powers of the charge, i.e. of SU(3) generators, are conceivable but are 
not considered here. Their contribution is suppressed by factors 
of $e^2/4\pi=1/137$. 

{\bf Rule 2:} With respect to SU(2)$_J$, electromagnetic multipoles transform 
according to their spatial tensor rank $J$ as discussed in 
sect.~\ref{sec:multop}. 
For example, quadrupole moments transform as rank $J=2$ tensors.
Consequently, in spin-flavor space, quadrupole moments transform according 
to the $({\bf 8},{\bf 5})$ product representation. The latter appears 
only in the SU(6) irreps ${\bf 405}$ and ${\bf 2695}$, 
which means that quadrupole moment operators must be constructed 
from two-quark and three-quark operators.

The second reason is that the spin-flavor decomposition of the SU(6) 
multiplets ${\bf 35}$, 
${\bf 405}$, and ${\bf 2695}$ shows which observables are connected 
by the underlying SU(6) symmetry. The relative weights of the different spin-flavor channels within a certain SU(6) representation are given by SU(6) Clebsch-Gordan
coefficients and are therefore exactly known irrespective of the fact that SU(6)
symmetry is broken. Applied to electromagnetic multipoles this means that
broken SU(6) symmetry relates multipoles of different tensor rank $J$. 
For example, the matrix elements of the charge monopole $({\bf 8},{\bf 1})$ 
and the charge quadrupole $({\bf 8},{\bf 5})$ operators are related, 
because  they belong to the same ${\bf 405}$ multiplet of SU(6). 
This provides the group-theoretical foundation of the relation 
in Eq.(\ref{qmnr}) and its generalization to finite 
momentum transfers as discussed in more detail in Appendix A.

\subsection{General spin-flavor parametrization of observables }
\label{sec:gp}
An efficient way to make use of the predictive power of broken SU(6) spin-flavor symmetry is the general parameterization (GP) method, developed by Morpurgo~\cite{Mor89,Mor89b}. The method is based on the symmetries and dynamics of QCD. Although 
noncovariant in appearance, all spin-flavor invariants that are allowed 
by Lorentz invariance and inner flavor symmetry are included in the operator basis.

The basic idea of this method is to {\it formally} define, 
for the observable at hand, a QCD operator ${\hat O}$ and QCD baryon eigenstates 
$\vert B \rangle$ expressed explicitly in terms of quarks and gluons. 
With the help of the unitary operator $V$, the original QCD matrix elements 
can be rewritten in a basis of auxiliary states $\vert\Phi_B \rangle $,
which are pure three-quark states with orbital angular momentum
$L=0$ and spin-flavor wave functions~\cite{Lic78} denoted as 
$\vert W_B\rangle $, that is 
\begin{equation}
\label{map}
\left \langle B \vert {\hat O} \vert B \right \rangle =
\left \langle \Phi_B \vert
V^{\dagger}{\hat O}  V \vert \Phi_B \right \rangle =
\left \langle W_B \vert
\mathcal{O} \vert W_B \right \rangle \, . 
\end{equation}   
The operator $V$ dresses the pure three-quark 
states $\vert \Phi_B \rangle$ with $q\overline q$ components and gluons and 
thereby generates
the exact QCD eigenstates $\vert B \rangle $ as in
\bea
\label{QCDstates}
\vert B\rangle & = &\alpha \vert qqq\rangle +\beta_1 \vert qqq\,(q\overline{q})\rangle 
+ \beta_2 \vert qqq\,(q\overline{q})^2\rangle + \ldots \nonumber \\
&+& \gamma_1 \vert qqq \, g\rangle + \gamma_2 \vert qqq \, gg\rangle + \ldots
\eea

On the right hand side of the last equality in Eq.(\ref{map}) 
the integration over spatial and color degrees of freedom 
has been performed. As a result only a matrix element 
of a spin-flavor operator $\mathcal{O}$ between spin-flavor states 
$\vert W_B\rangle$ remains.
The spatial and color matrix elements
are absorbed into {\it a priori} unknown parameters multiplying 
the spin-flavor invariants appearing in the expansion of the operator 
$\mathcal{O}$. The eliminated quark-antiquark and gluon degrees of freedom 
are effectively described by symmetry breaking many-quark operators
in spin-flavor space~\cite{Mor89,Mor89b}. 

A general expression of the spin-flavor operator $\mathcal{O}$ for a given observable can then be constructed as a sum of one-, two-, and three-quark operators
\be
\label{onetwothree} 
\mathcal{O} = \mathcal{O}_{[1]} + \mathcal{O}_{[2]} + \mathcal {O}_{[3]},
\ee
which transform according to the ${\bf 35}$, ${\bf 405}$ and ${\bf 2695}$ 
dimensional representations of SU(6) respectively as given in Eq.(\ref{baryon56}). 
%%%%%%%%%%%%%%%%%%%%%%%%%%%%%%%%%%%%%%%%%%%%%%%%%%%%%%%%%%%%%%%%%%%%%%%%%
%             FIG. 4:  One-body, two-body, three-body currents
%%%%%%%%%%%%%%%%%%%%%%%%%%%%%%%%%%%%%%%%%%%%%%%%%%%%%%%%%%%%%%%%%%%%%%%%%
\begin{figure*}
\begin{center}
% Use the relevant command to insert your figure file.
% For example, with the graphicx package use
  \includegraphics[width=0.80\textwidth]{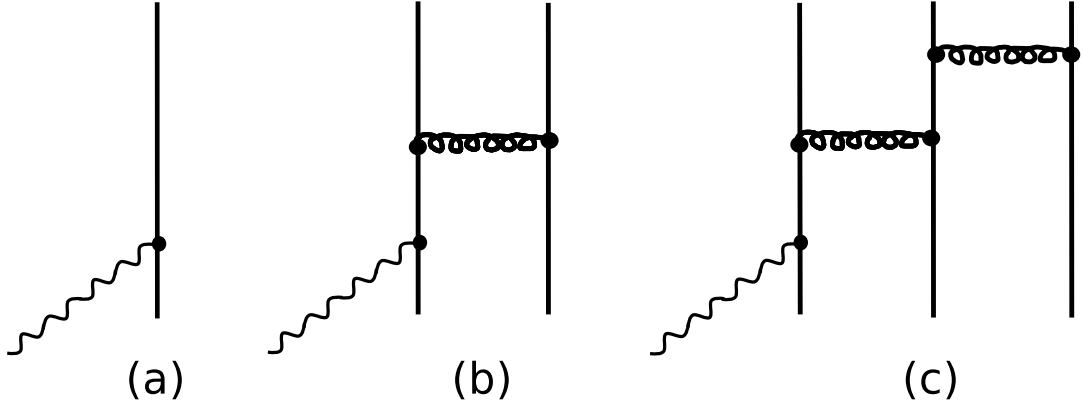}
% figure caption is below the figure
\caption{Fundamental photon-quark processes contributing to 
electromagnetic multipole form factors in Fig.~\ref{fig:scattering}:
(a) one-quark current ($\rho_{[1]},{\bf J}_{[1]}$), 
(b) two-quark current ($\rho_{[2]},{\bf J}_{[2]}$), 
(c) three-quark current ($\rho_{[3]},{\bf J}_{[3]}$).}
\label{fig:one_two_three}       % Give a unique label
\end{center}
\end{figure*}
For electromagnetic currents the physical interpretation of these operator structures 
in Eq.(\ref{onetwothree}) is as follows. 
The one-quark operator ${\mathcal O}_{[1]}$ in Fig.~\ref{fig:one_two_three}(a) 
can be interpreted as the valence 
quark contribution, whereas the two-quark term ${\mathcal O}_{[2]}$ 
and the three-quark term ${\mathcal O}_{[3]}$ reflect the $q \overline q$ 
and gluon degrees of freedom that have been eliminated from the Hilbert space
spanned by the QCD baryon states in Eq.(\ref{QCDstates}).
For example, the two-quark operator constructed from Fig.~\ref{fig:one_two_three}(b), 
reflects quark-antiquark and gluon degrees of freedom. This becomes apparent after projecting the covariant quark propagator between photon absorption and gluon emission 
onto the negative intermediate energy component of the propagator~\cite{Buc90}. 

The GP method has been used to calculate various baryon 
properties~\cite{Buc08,Mor89,Mor89b,Mor99,Hen00,Hen02,Buc11,Buc14}. 
As a rule one finds that one-quark operators are more important than 
two-quark operators, 
which in turn are more important than three-quark operators. There are however some
important exceptions to this rule. 
If one-quark operators give a vanishing contribution (neutral baryon charge radii)
or are forbidden due to selection rules (quadrupole moments), 
two-quark operators dominate. 
Similarly, if, as in the case of octupole moments, one- and two-quark operators are 
forbidden, three-quark operators provide the dominant contribution.
 
The SU(6) symmetry analysis and GP method are connected with the underlying 
field theory of QCD. This is becomes apparent in the $1/N_c$ expansion of QCD processes. 

\subsection{The $1/N_c$ expansion of QCD observables}
\begin{figure*}
\begin{center}
\includegraphics[width=0.75\textwidth]{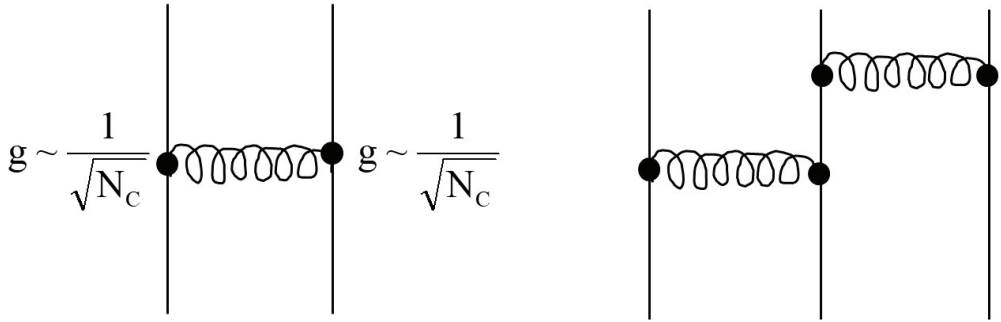}
\caption{The strong coupling $\alpha_S(Q^2)=g^2(Q^2)/(4\pi)=
{12\pi}\,[(11 N_c-2 N_f) \ln(Q^2/\Lambda^2)]^{-1} \sim 1/N_c$ is inversely
proportional to the number of colors $N_c$.
Multigluon exchange diagrams involve higher powers of $g \sim 1/\sqrt{N_c}$.
Therefore, two-quark operators are typically suppressed by $1/N_c$ and 
three-quark operators by $1/N_c^2$ compared to one-quark operators.}
\label{fig:largen}       % Give a unique label
\end{center}
\end{figure*}

The seminal work on calculating baryon observables using the $1/N_c$ expansion
is by Witten~\cite{wit79}. Later the relation between QCD and broken spin-flavor 
symmetry underlying the parametrization method was made apparent 
in the limit $N_c \to \infty$, in which case the QCD Lagrangian has an
exact spin-flavor symmetry~\cite{das94}.
For finite $N_c$, spin-flavor symmetry is broken but the method allows to 
classify spin-flavor symmetry breaking operators 
according to the powers of $1/N_c$ associated with them.
It turns out that second and third higher order SU(6) symmetry breaking operators 
${\mathcal O}_{[2]}$ and ${\mathcal O}_{[3]}$ are 
suppressed by $1/N_c$ and $1/N_c^2$ respectively, compared
to the first order symmetry breaking one-quark operators ${\mathcal O}_{[1]}$ 
thus explaining the hierarchy observed in the GP method. This is qualitatively illustrated in Fig.~\ref{fig:largen}.
For a review see Ref.~\cite{leb99}.

The $1/N_c$ expansion method has been applied to a number of observables 
in particular to baryon charge radii and quadrupole moments~\cite{Hes02,Leb00},
and numerous relations among baryon charge radii and quadrupole moments have been
found.
%%%%%%%%%%%%%%%%%%%%%%%%%%%%%%%%%%%%%%%%%%%%%%%%%%%%%%%%%%%%%%%%%%%%%%%%%
%             FIG. 5:  1/N behavior of quark operators
%%%%%%%%%%%%%%%%%%%%%%%%%%%%%%%%%%%%%%%%%%%%%%%%%%%%%%%%%%%%%%%%%%%%%%%%%
For example the relation between the neutron charge radius $r_n^2$ 
and $N\to \Delta$ transition quadrupole moment $Q_{N \to \Delta}$ in Eq.(\ref{qmnr})
has been investigated using the $1/N_c$ expansion method~\cite{Hes02}. 
Including second and third order SU(6) symmetry breaking operators
the following expression has been found:
\be
\label{ffrel2statlargeN}
Q_{N \to \Delta}= \frac{1}{\sqrt{2}} \, r_n^2 \, \, 
\left ( \frac{N_c}{N_c +3} \, \sqrt{\frac{N_c+5}{N_c-1}} \right ). 
\ee
It is interesting that this more general relation 
is equivalent to Eq.(\ref{qmnr}) both for the physical $N_c=3$ case 
and for $N_c \to \infty$. For arbitrary $N_c$, the difference 
between Eq.(\ref{ffrel2statlargeN}) and Eq.(\ref{qmnr}) is always less than 1.2$\%$.

Up to now we have discussed the application of 
three different SU(6) symmetry based methods 
to the ${\bf 56}$ dimensional representation of ground state baryons 
for which $J=S$ and $L=0$.
When the orbital angular momentum $L$ of the states and operators is nonzero
as for the $N^*(1680)$ resonance, 
the symmetry group has to be enlarged to 
SU(6)$\times$O(3). This was done in the case of the $1/N_c$ expansion 
by several authors~\cite{Goi07,Coh05,Mat16}. In this work we use a fourth group theoretical method, namely current algebra. In sect.~\ref{subsec:current_algebra}
we employ current algebra to calculate transition multipole moments of excited states with $L\neq 0$. Current algebra stresses the importance of the commutation
relations between group generators and explores the consequences 
that follow from this symmetry requirement.

\subsection{Algebra of vector and axial vector current components}
\label{subsec:current_algebra}
The algebra of electromagnetic and weak currents provides a group theoretical 
description of the structure of hadrons 
based on the concept that the vector and axial vector currents are proportional to 
group generators. Clearly, the electromagnetic currents 
involve the SU(3) generators 
$T_3=\lambda_3/2$ (isovector current $J^3$) and $Y=\lambda_8/\sqrt{3}$ 
(isoscalar current $J^8$) occuring in the Gell-Mann-Nishijima
relation of Eq.(\ref{GMNR}) for the electric charge ${\rm{Q}}$.
To describe weak vector currents, the isovector $T_3$ term of the 
electromagnetic current in Eq.(\ref{GMNR}) is generalized to strangeness conserving 
weak isovector-vector currents based on $T_{\pm}=(\lambda_1 \pm i \lambda_2)/2$, 
in addition to strangeness changing weak vector currents involving 
$V_{\pm}=(\lambda_4 \pm i \lambda_5)/2$ and
$U_{\pm}=(\lambda_6 \pm i \lambda_7)/2$. 
The space integrals of these 8 vector currents $J^{\gamma}$ with $\gamma=1,...,8$ obey the same SU(3) commutation relations (Lie algebra) as the SU(3) generators~\cite{Gel64},
\be
\label{commrel}
\left [ \lambda_{\alpha}, \lambda_{\beta} \right ] =
 2 i f_{\alpha \beta \gamma} \lambda_{\gamma},
\ee 
where the $f_{\alpha \beta \gamma}$ are the antisymmetric SU(3)$_F$ structure constants.

An analogous SU(3)$_A$ algebra describes weak axial currents $J^{5\gamma}$.
Thus, in electroweak theory one is dealing with a 
SU(3)$_V \times$SU(3)$_A$ group.
When taking the linear combinations 
\be
J^{\gamma \pm} = J^{\gamma } \pm  J^{5 \,  \gamma},
\ee
the new generators $J^{\gamma \pm}$ satisfy a closed system of 
commutation relations~\cite{Gel64}. 
As emphasized by Gell-Mann, no matter how badly SU(3) 
flavor symmetry is broken, the group generators satisfy the algebraic commutation relations exactly. This observation is the basis 
of several important sum rules, such as the Adler-Weisberger sum rule 
relating the weak axial coupling to pion-nucleon scattering cross sections~\cite{deA73}.

A further generalization based on the relativistic vector and axial-vector quark flavor currents involves the Dirac matrices $\gamma_{\mu}$ with $\mu=0,...3$ and $\gamma_5$
\be 
J_{\mu}^{\alpha}=\bar{q} \,\frac{1}{2}\lambda_{\alpha} \, \gamma_{\mu}\, q 
\ \ \ J_{\mu}^{5, \alpha}=\bar{q}\, \frac{1}{2}\, \lambda_{\alpha} 
\, \gamma_{\mu} \gamma_5\,  q,
\ee
where $\alpha=0,...,8$ and $\lambda_0=\sqrt{2/3}\,\, {\bf 1}$. 
This leads to an algebra of $72=8 \cdot 4 \cdot 2$ vector and axial current components corresponding to the generators of a chiral U(6)$_V\times$U(6)$_A$ algebra.
In the following, we use this generalized form of Gell-Mann's current algebra~\cite{Fey64} in which the time and spatial components of the vector current densities satisfy the following commutation relations~\cite{Das65,Lee65,Bie66}
\bea
\label{current_algebra}
%charge-charge commutator
\left [J_0^{\alpha}({\bf r}), \, J_0^{\beta}({\bf r}') \right ] & = &i\, 
f_{\alpha \beta \gamma}
 \delta({\bf r} - {\bf r}') \, J_0^{\gamma}({\bf r}), \nonumber \\
%mixed charge-current commutator
%\left [J_0^{\alpha}({\bf r}), \, J_i^{\beta}({\bf r}') \right ] & = &i\, 
%f_{\alpha \beta \gamma}\,
% \delta({\bf r} - {\bf r}') \, J_i^{\gamma}({\bf r}) 
%\nonumber \\
%current-current commutator
\left [J_i^{\alpha}({\bf r}), \, J_j^{\beta}({\bf r}') \right ] & = &i\, 
f_{\alpha \beta \gamma}
\delta_{ij}  \delta({\bf r} - {\bf r}') \, J_0^{\gamma}({\bf r}) 
\nonumber \\
 &+ &  i  d_{\alpha \beta \gamma'}
\epsilon_{ijk}  \delta({\bf r} - {\bf r}') \, A_k^{\gamma'}({\bf r}).
\eea
The flavor components of the spatial current $J_i^{\alpha}$ and 
charge $J_0^{\alpha}$ densities  are denoted by greek superscripts 
$\alpha, \beta, \gamma$. The roman subscripts $i,j,k$ 
indicate the cartesian components of the 
spatial vector ${\bf J}$ and axial vector ${\bf A}={\bf J}^5$ currents. 
As usual, $\delta_{ij}$ and $\epsilon_{ijk}$ refer to the Kronecker and 
Levi-Civita tensors, and the $d_{\alpha \beta \gamma}$ are the symmetric
SU(3) structure constants. 

An early application of the current algebra 
method to magnetic moments led to
the Gell-Mann Dashen relation between the proton magnetic moment $\mu_p$ and 
the proton charge radius $r_p$~\cite{Das65}
\be 
\label{GMDR}
\mu_p^2 = \frac{1}{6} r_p^2, 
\ee
where $\mu_p$ is expressed in nuclear magnetons $\mu_N=1/(2 M_N)$ in units [fm]. Eq.(\ref{GMDR}) is satisfied within 20$\%$.
In our application to quadrupole and octupole transition multipole moments 
we will also take space integrals 
of these charge current components similar to the work of Bietti~\cite{Bie66}. 

\section{Results}

\subsection{Charge radii of ground state baryons}
\label{subsec:charad}
As in Eq.(\ref{cha_mult_exp}) we expand the baryon charge density 
operator $\rho({\bf q})$ 
into Coulomb multipoles with projection $M=0$ up to quadrupole terms 
\bea
 \rho(q)   & \! \! \! = \! \! \!   & \sqrt{4 \pi} 
\sum_J i^J {\hat J}  T_0^{CJ}(q) \! \!  = \! \! 
\rho^{C0}(q) + \rho^{C2}(q) + ...,
\eea
which have been evaluated for $\hat{{\bf q}}={\bf e}_z$ so that 
$Y^J_0({\hat{\bf q}})= {\hat J}/\sqrt{4 \pi}$
with ${\hat J}=\sqrt{2J+1}$. The lowest moments of 
$\rho$ are then obtained from a low momentum transfer expansion of
$j_J(gr)$ in Eq.(\ref{cmo}). Up to $q^2$ contributions one has 
\be
\rho(q) = {\rm Q} - \frac{q^2}{6} r^2 -\frac{q^2}{6} {\mathcal Q} + ...
\ee
The first two terms arise from the spherically symmetric monopole 
$\rho^{C0}$
part and the third term comes from the quadrupole $\rho^{C2}$
part of $\rho$. The low $q$ expansion of $\rho$ 
gives the baryon's total charge (${\rm Q}$),
spatial extension ($r^2$), and shape (${\mathcal Q}$).

According to the group theoretical approach outlined in sect.~\ref{sec:spin-flavor} and
sect.~\ref{sec:gp}, the charge radius is a rank $J=0$ operator and must be constructed 
as a sum of one-, two-, and three-quark terms,  
each of which transforming as an ${\bf (8,1)}$ representation 
in flavor-spin space, i.e. as a flavor octet and a spin scalar 
\begin{equation}
\label{para1}
{r}^2 =  A \sum_{i=1}^3 e_i {\bf 1} + 
B\sum_{i \ne j}^3 e_i \, \xbf{\sigma}_i \cdot \xbf{\sigma}_j  + 
C\!\! \sum_{i \ne j \ne k }^3 e_k \, \xbf{\sigma}_i \cdot \xbf{\sigma}_j, 
\end{equation}
where
$e_i=(1 + 3 \tau_{i \, z})/6$ and $\xbf{\sigma}_i$ are 
the charge and spin operators of the i-th quark.
Here, $\tau_{i \, z}$ denotes the $z$ component of the Pauli isospin matrix.
These are the only allowed spin scalars and flavor octets 
that can be constructed from the 
generators of the spin-flavor group in Eq.(\ref{SU(6)_generators}). 
The constants $A$, $B$, and $C$ 
parametrizing the orbital and color matrix elements 
are determined from experiment.

Nucleon and $\Delta$ charge radii are then calculated by evaluating 
matrix elements of the operator in Eq.(\ref{para1}) 
between three-quark spin-flavor wave functions $\vert W_B \rangle$
\be
\label{charadmm}
r_B^2 = \langle W_B \vert {r}^2 \vert W_B \rangle .
\ee
For charged baryons, $r^2_B$ is normalized by dividing by the baryon charge. 
The results for octet and decuplet baryons are summarized in 
Table~\ref{tab:baryonradii}. A complete table including all 18 ground 
state baryon charge radii and the relations between them is given in
Ref.~\cite{Buc07}.   The results agree with those in Ref.~\cite{Leb00} after setting $N_c=3$ and an obvious redefinition of the constants.
%%%%%%%%%%%%%%%%%%%%%%%%%%%%%%%%%%%%%%%%%%%%%%%%%%%%%%%%%%%%%%%%%%%%%%%%%
% octet and decuplet  charge radii
%%%%%%%%%%%%%%%%%%%%%%%%%%%%%%%%%%%%%%%%%%%%%%%%%%%%%%%%%%%%%%%%%%%%%%%%
\begin{table}[t]
\begin{center}
\caption[chargeradii]{Nucleon and $\Delta$ charge radii  
in [fm$^2$] with one-quark $(A)$, two-quark ($B$), 
and three-quark ($C$) contributions. 
Left: Analytic expressions for $r^2_B$ obtained from Eq.(\ref{charadmm}). 
Right: Numerical values using $r_p=0.8751(61)$ fm~\cite{Pat16}, 
$r^2_n=-0.1149(35)$ fm$^2$~\cite{Kop97},
$r^2_{{\Sigma}^-}=0.61(12)$ fm$^2$~\cite{Esc01} as input yielding $A=0.7299$, 
$B=0.0455$, and $C=-0.0060$ in [fm$^2$]. For details see Ref.~\cite{Buc07}.}
\label{tab:baryonradii}
\begin{tabular}{| l | l | r |} 
\hline
                &   $\ \ \ \  \ \ r_B^2$             & $r_B^2$  [fm$^2$]   \\ 
\hline
$n$             & $\phantom{A}-2B+4C$   &    -0.115 \\
%%%%%%%%%%%%%%%%%%%%%%%%%%%%%%%%%%%%%%%%%%%%%%%%%%%%%%%%%%%%%%%%%%%%%%%%%%
$p$             & $A \phantom{\,\,2B+}-6C$ &  0.766 \\

%%%%%%%%%%%%%%%%%%%%%%%%%%%%%%%%%%%%%%%%%%%%%%%%%%%%%%%%%%%%%%%%%%%%%%%%%%
\hline
\hline
%%%%%%%%%%%%%%%%%%%%%%%%%%%%%%%%%%%%%%%%%%%%%%%%%%%%%%%%%%%%%%%%%%%%%%%%%%
$\Delta^-$      & $A +2B +2C$ & 0.809 \\
%%%%%%%%%%%%%%%%%%%%%%%%%%%%%%%%%%%%%%%%%%%%%%%%%%%%%%%%%%%%%%%%%%%%%%%%%%
$\Delta^{0}$    & $\phantom{+++++}0$ & 0.809 \\
%%%%%%%%%%%%%%%%%%%%%%%%%%%%%%%%%%%%%%%%%%%%%%%%%%%%%%%%%%%%%%%%%%%%%%%%%%
$\Delta^{+}$    & $A +2B +2C $  & 0.809 \\
%%%%%%%%%%%%%%%%%%%%%%%%%%%%%%%%%%%%%%%%%%%%%%%%%%%%%%%%%%%%%%%%%%%%%%%%%%
$\Delta^{++}$   & $A +2B +2C $ & 0.809\\
%%%%%%%%%%%%%%%%%%%%%%%%%%%%%%%%%%%%%%%%%%%%%%%%%%%%%%%%%%%%%%%%%%%%%%%%%% 
\hline
\end{tabular}
\end{center}
\end{table}

\subsection{Quadrupole moments of ground state baryons}
\label{subsec:quadmom}
As explained in sect.~\ref{sec:multisym} the charge quadrupole operator is constructed 
from flavor ${\bf 8}$ and spin $J=2$ operators as a sum of two- and three-body quark 
terms each transforming as an ${\bf (8,5)}$ representation in flavor-spin space
\bea
\label{para2}
\mathcal{Q}  & = & B'\sum_{i \ne j}^3 e_i 
\left ( 3 \sigma_{i \, z} \sigma_{ j \, z} -
\xbf{\sigma}_i \cdot \xbf{\sigma}_j \right ) \nonumber \\ 
& + & C'\!\!\sum_{i \ne j \ne k }^3 e_k 
\left ( 3 \sigma_{i \, z} \sigma_{ j \, z} -
\xbf{\sigma}_i \cdot \xbf{\sigma}_j \right ). 
\eea
Baryon decuplet quadrupole moments $Q_{B^*}$ and octet-decu\-plet tran\-sition
qua\-dru\-pole mo\-ments $Q_{B \to B^*}$ are obtained by calculating the
matrix elements of the quadrupole operator
in Eq.(\ref{para2}) between the three-quark spin-flavor wave 
functions $\vert W_B \rangle$ and $\vert W_{B^*} \rangle$
\be
\label{quadmom_matrixelements}
Q_{B^*}  \!\!=\!\! \left \langle W_{B^*} \vert \mathcal{Q}
\vert W_{B^*} \right \rangle, \, \, 
Q_{B \to B^*}
\!\! =  \!\! \left \langle W_{B^*} \vert \mathcal{Q} \vert W_B \right \rangle,
\ee
where $B$  denotes a spin 1/2 octet baryon and $B^*$ a member of the
spin 3/2 baryon decuplet.  The ensuing results for quadrupole moments and the relations
between them have been discussed earlier~\cite{Hes02,Hen02,Leb00}. 
Table~\ref{tab:quadmo} reproduces some pertinent results.
%%%%%%%%%%%%%%%%%%%%%%%%%%%%%%%%%%%%%%%%%%%%%%%%%%%%%%%%%%%%%%%%%%%%%%
\begin{table}[b]
\begin{center}
\caption[C2 moments]{Transition and diagonal baryon quadrupole moments
with two-quark (B') and three-quark (C') contributions~\cite{Hen02,Buc07}.
Left: Analytic expressions for $Q_B$ obtained from Eq.(\ref{quadmom_matrixelements}). 
Right: Numerical values using the parameter set of Table~\ref{tab:baryonradii}
with $B'=-B/2$, $C'=-C/2$, and $\zeta=m_u/m_s=0.613$.
We then obtain $Q_{p \to \Delta^+}({\rm theory})=-0.0812(25)$ fm$^2$ 
with $r_n^2$ as input compared to 
$Q_{p \to \Delta^+}({\rm exptl}) = -0.0846(33)$ fm$^2$~\cite{Tia11} and 
$Q_{p \to \Delta^+}({\rm exptl}) = -0.108(9)$ fm$^2$~\cite{Bla01}.
}
\label{tab:quadmo}
\begin{tabular}{|l | c | r |}
\hline
      &     $\ \ \ \ Q_B$            & $Q_B$ [fm$^2$]    \\ 
\hline
%%%%%%%%%%%%%%%%%%%%%%%%%%%%%%%%%%%%%%%%%%%%%%%%%%%%%%%%%%%%%%%%%%%%%%%%
$n\to \Delta^0$  &  $2\sqrt{2}(B'-2C')$   & -0.081    \\
%%%%%%%%%%%%%%%%%%%%%%%%%%%%%%%%%%%%%%%%%%%%%%%%%%%%%%%%%%%%%%%%%%%%%%%%
$p\to \Delta^+$   & $2\sqrt{2}(B'-2C')$  & -0.081       \\
%%%%%%%%%%%%%%%%%%%%%%%%%%%%%%%%%%%%%%%%%%%%%%%%%%%%%%%%%%%%%%%%%%%%%%%%
\hline
\hline
$\Delta^{-}$      & $ -4B' -4C'$  &  0.079  \\
%%%%%%%%%%%%%%%%%%%%%%%%%%%%%%%%%%%%%%%%%%%%%%%%%%%%%%%%%%%%%%%%%%%%%%%%
$\Delta^{0}$      &     0        &   0 \\
%%%%%%%%%%%%%%%%%%%%%%%%%%%%%%%%%%%%%%%%%%%%%%%%%%%%%%%%%%%%%%%%%%%%%%%%
$\Delta^{+}$      & $4B' +  4C'$ &  -0.079  \\
%%%%%%%%%%%%%%%%%%%%%%%%%%%%%%%%%%%%%%%%%%%%%%%%%%%%%%%%%%%%%%%%%%%%%%%%
$\Delta^{++}$     & $8B' + 8C'$   & -0.158   \\
%%%%%%%%%%%%%%%%%%%%%%%%%%%%%%%%%%%%%%%%%%%%%%%%%%%%%%%%%%%%%%%%%%%%%%%%
$\Omega^-$        & $-(4B' + 4C')\zeta^3 $  & 0.018        \\
%%%%%%%%%%%%%%%%%%%%%%%%%%%%%%%%%%%%%%%%%%%%%%%%%%%%%%%%%%%%%%%%%%%%%%%% 
\hline
\end{tabular}
\end{center}
\end{table}
%%%%%%%%%%%%%%%%%%%%%%%%%%%%%%%%%%%%%%%%%%%%%%%%%%%%%%%%%%%%%%%%%%%%%%

In this work, we are mainly concerned with relations between the measureable 
transition quadrupole moments and nucleon ground state properties. To this end, we
will first show that $B'=-B/2$. One can understand the result $B'=-B/2$ 
from the explicit 
expression~\cite{Buc91} for the one-gluon exchange charge density 
in Fig.~\ref{fig:one_two_three} 
\bea
\label{gluexcur1}
\rho_{gq{\bar q}} &= &-i\frac{\alpha_S}{16 m_q^3} 
\sum_{i < j}^3 \xbf{\lambda}_i \cdot \xbf{\lambda}_j 
\bigl( e_i e^{i{\bf q} \cdot {\bf r}_i} 
(\xbf{\sigma}_i \times {\bf q} ) \cdot (\xbf{\sigma}_j \times {\bf r} )
\nonumber \\
& +  & (i \leftrightarrow j) \bigr ) \frac{1}{r^3},
\eea
where $\xbf{\lambda}_i$ are the SU(3) color matrices of quark i. Here we have reproduced 
only the spin-dependent terms of $\rho_{gq{\bar q}}$ that contribute to $r_n^2$ 
and $Q_{p \to \Delta^+}$.
%%%%%%%%%%%%%%%%%%%%%%%%%%%%%%%%%%%%%%%%%%%%%%%%%%%%%%%%%%%%%%%%%%%%%%%%%
%             FIG. 6:  Double spin-flip N-->Delta quadrupole transition
%%%%%%%%%%%%%%%%%%%%%%%%%%%%%%%%%%%%%%%%%%%%%%%%%%%%%%%%%%%%%%%%%%%%%%%%%
\begin{figure*}
\begin{center}
% Use the relevant command to insert your figure file.
% For example, with the graphicx package use
  \includegraphics[width=0.75\textwidth]{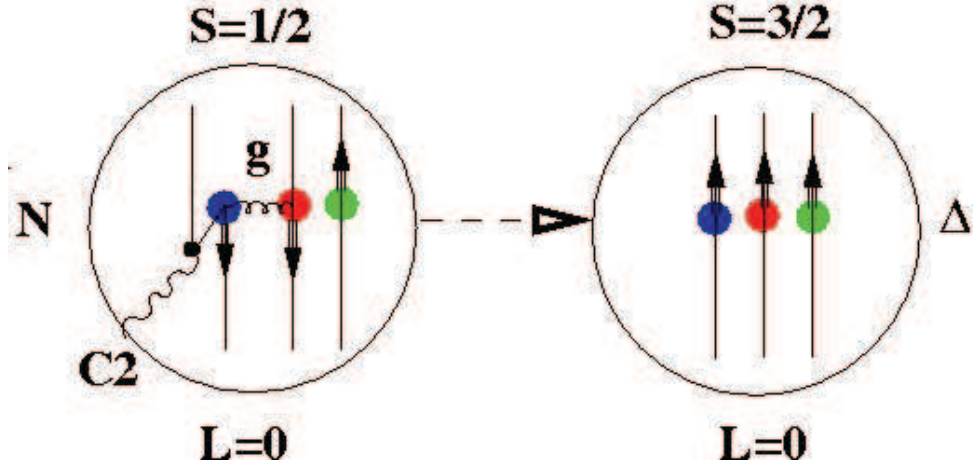}
% figure caption is below the figure
\caption{The $q{\bar q}$-gluon exchange charge density $\rho_{[2]}$ in Fig.~\ref{fig:one_two_three} 
induces the $N \to \Delta$ charge quadrupole (C2) transition 
via a double spin-flip of two quarks, i.e. via the spin tensor term in 
Eq.(\ref{gluexcur2}). Siegert's theorem connects this 
double spin-flip process with a spatial exchange current providing
the dominant contribution to the $N\to \Delta$ transverse electric (E2) 
form factor~\cite{Buc98}.}
\label{fig:doublespinflip}   % Give a unique label
\end{center}
\end{figure*}
%%%%%%%%%%%%%%%%%%%%%%%%%%%%%%%%%%%%%%%%%%%%%%%%%%%%%%%%%%%%%%%%%%%%%%%%%
After some angular momentum recoupling we can rewrite Eq.(\ref{gluexcur1})
as a superposition of a spin scalar and a spin tensor term 
depicted in Fig.~\ref{fig:doublespinflip} with definite relative weight as 
\be
\label{gluexcur2}
\rho_{gq\overline{q}} =
B\sum_{i \ne j}^3 e_i \left[ \,(-2)\, \xbf{\sigma}_i \cdot \xbf{\sigma}_j 
+(3 \sigma_{i \, z} \sigma_{ j \, z} -\xbf{\sigma}_i \cdot \xbf{\sigma}_j) \right]. 
\ee
The factor $B$ contains the radial, momentum, and color dependence common to both
spin-dependent terms. Thus, for the gluon exchange charge density shown in 
Fig.~\ref{fig:one_two_three} there is a fixed ratio of (-2) between the spin scalar and spin tensor parts of the corresponding operator. The same relative factor is obtained 
for pion exchange or a combination of gluon and pion exchange between quarks. 
The relative factor (-2) between spin scalar and spin tensor turns out to be 
a model-independent symmetry based property of two-quark charge densities.

If the fixed ratio between spin scalar and spin tensor is implemented 
in the GP method, relation Eq.(\ref{qmnr}) follows from the expressions 
in Table~\ref{tab:baryonradii} and Table~\ref{tab:quadmo} as
\be
\label{qmnr2}
Q_{p \to \Delta^+} = 2\sqrt{2} (B'-2C')= -\sqrt{2} (B-2C) =\frac{1}{\sqrt{2}}\, r_n^2.
\ee
In appendix A, we provide a group theoretical derivation of this fixed ratio and of
Eq.(\ref{qmnr}) based on broken spin-flavor symmetry without making any dynamical assumptions.

%%%%%%%%%%%%%%%%%%%%%%%%%%%%%%%%%%%%%%%%%%%%%%%
%             FIG. 7: ratio C2/M1 Maid 2011
%%%%%%%%%%%%%%%%%%%%%%%%%%%%%%%%%%%%%%%%%%%%%%%
\begin{figure*}
\begin{center}
\includegraphics[height=0.40\textheight]{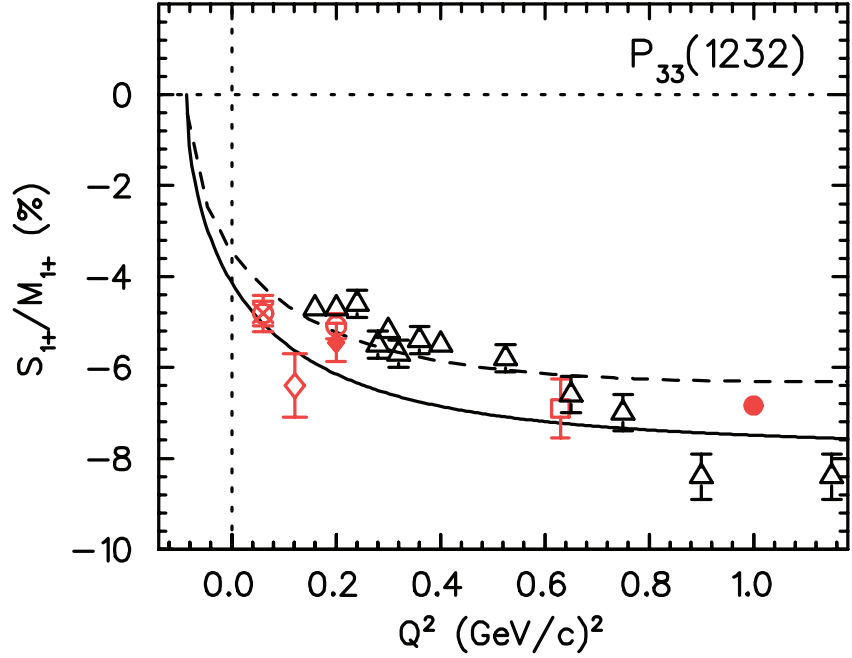}
\caption{
\label{C2_m1maid07}
The $C2/M1(Q^2)\equiv S_{1+}/M_{1+}(Q^2)$ ratio for 
low four-momentum transfers. 
The full curve is a fit of the experimental $C2/M1$ ratio as determined 
from the world electro-pionproduction data.
The dashed curve is calculated using the form factor relation of 
Eq.(\ref{C2C0rel}). Figure taken from Ref.~\cite{tia16}.}
\end{center}
\end{figure*}

From the form factor relation in Eq.(\ref{C2C0rel}) we can also extract 
the quadrupole transition radius, which is determined by the fourth and 
second radial moments of the neutron charge distribution~\cite{Buc09}
\be
r^2_{C2} = \frac{7}{10}\, \frac{r^4_n}{r^2_n}.
\ee
From the radial moments of the neutron charge density
in Table~\ref{tab:higherradmom} we find $r_{C2} = 1.68 \, {\rm fm}$. 
Thus, $r_{C2} \approx r_{\pi}$ is close to the pion Compton wavelength. 
We have previously suggested that $r_{C2}$
measures the spatial extension of the $q{\bar q}$ pair distribution 
in the nucleon~\cite{Buc09}.  Recently, $r_{C2}$ has been determined from 
combined fits of the $G_C^n$ and $G_{C2}^{N \to \Delta}$ form factor 
data~\cite{Ram17} obtaining a somewhat smaller value $r_{C2} = 1.32$ fm.

Broken SU(6) spin-flavor symmetry also leads to the following
relation~\cite{Beg64} between the neutron elastic and the $N \to \Delta$ transition 
magnetic form factors, and at $Q^2=0$ between the neutron and $N \to \Delta$ transition 
magnetic moments 
\be 
\label{M1rel}
G_{M1}^{N \to \Delta}(Q^2)  =  -\sqrt{2}\, G_M^n(Q^2), 
\quad \mu_{N \to \Delta}= -\sqrt{2} \, \mu_{n}.
\ee
With the help of Eq.(\ref{C2C0rel}) and the magnetic form factor relation 
of Eq.(\ref{M1rel}), the C2/M1 ratio in electromagnetic $\Delta(1232)$ 
excitation can be expressed in terms of the neutron elastic form factors as 
follows~\cite{Buc04} 
\bea
\label{C2M1ratio}
\frac{C2}{M1}(Q^2) & := & \frac{\vert {\bf q} \vert \, M_N}{6}  
\, \frac{G_{C2}^{N \to \Delta}(Q^2)}{G_{M1}^{N \to \Delta}(Q^2)} =  \frac{ \vert {\bf q}\vert M_N}{2 \,Q^2} \, \frac{G_{C}^{n}(Q^2)}{G_M^n(Q^2)}, \nonumber \\ 
\frac{C2}{M1}(0) & = & -\frac{(M_{\Delta}^2-M_N^2)}{2 M_{\Delta}} \, \frac{M_N}{12}
\frac{r_n^2}{\mu_n}.
\eea
where $\vert {\bf q} \vert$ is the modulus of the photon's three-momentum 
and $M_N$, $M_{\Delta}$ are the nucleon and $\Delta$ masses. 
The dashed curve in Fig.~\ref{C2_m1maid07} shows that the prediction based 
on Eq.(\ref{C2M1ratio}) agrees quite well with the data. Moreover, it has the 
correct limiting behavior for $Q^2\to 0$ and $Q^2 \to \infty$~\cite{Idi04,tia16}.
In particular, for $Q^2=0$ we get $C2/M1=-3.1\%$ in good agreement with 
recent experimental results~\cite{Blo16}. 

The electromagnetic $N \to \Delta$ transition multipoles also affect other 
observables. It has recently been shown~\cite{Hag18} that the $\Delta$(1232) resonance 
has an appreciable impact on the spectrum of atomic hydrogen. 
%%%%%%%%%%%%%%%%%%%%%%%%%%%%%%%%%%%%%%%%%%%%%%%%%%%%%%%%%%%%%%%%%%%%%%%%%%%%%%%%%%%%

\subsection{Intrinsic charge quadrupole form factor of the nucleon}
\label{subsec:intrinsicqm}

To study the implications of Eq.(\ref{qmnr}) and Eq.(\ref{C2C0rel}) for the shape 
of the nucleon ground state it is important to distinguish between the 
{\it spectroscopic} and {\it intrinsic} quadrupole moment of a particle~\cite{Boh75}.
It is known that a vanishing spectroscopic quadrupole moment due to angular 
momentum selection rules does not necessarily imply a spherically symmetric charge
distribution. For deformed spin 0 and spin 1/2 nuclei this insight has led to the 
general concept 
of an intrinsic quadrupole moment, which can be defined for different nuclear 
models.  The notion of an intrinsic quadrupole moment allows us to interpret 
measurable transition quadrupole moments in terms of the shape of the ground state. 

\begin{figure*}
\begin{center}
% Use the relevant command to insert your figure file.
% For example, with the graphicx package use
  \includegraphics[width=0.75\textwidth]{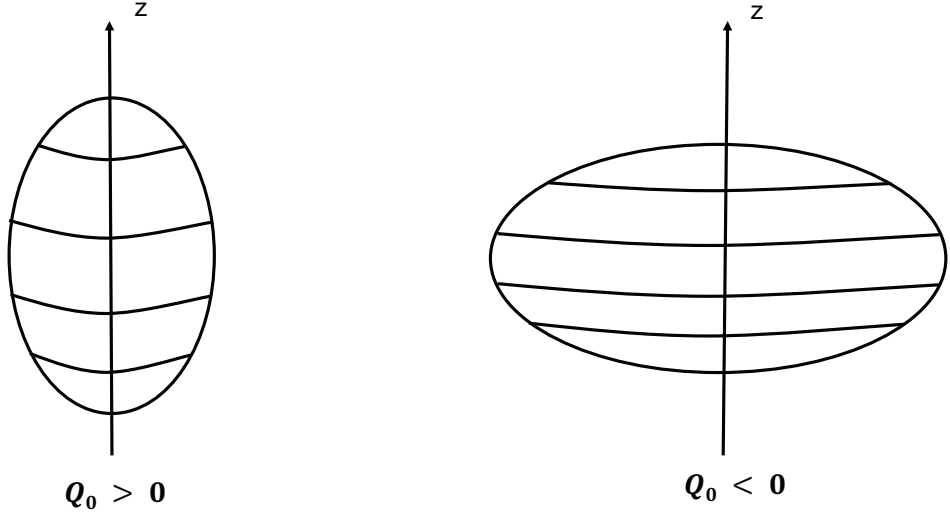}
% figure caption is below the figure
\caption{Prolate (left) and oblate (right) charge distribution corresponding to a positive (negative) intrinsic quadrupole moment $Q_0$ defined as 
in Eq.(\ref{intquadmom}). An analogous figure is obtained for the current distribution with $Q_0$ replaced by $\Omega_0$. }
\label{fig:prolate}       % Give a unique label
\end{center}
\end{figure*}

The geometric shape of a spatially extended particle is determined by 
its {\it intrinsic} quadrupole moment,
\be 
\label{intquadmom}
Q_0=\int \! \!d^3r \, \rho({\bf r}) \,  (3 z^2 - r^2), 
\ee
which is defined with respect to the body-fixed frame. 
If the charge density is concentrated along the $z$-direction 
(symmetry axis of the particle),  
the term proportional to $3z^2$ dominates, $Q_0$ 
is positive, and the particle is prolate (cigar-shaped).
If the charge density is concentrated in the equatorial plane perpendicular
to $z$, the term proportional to $r^2$ prevails, $Q_0$
is negative, and the particle is oblate (pancake-shaped) as depicted in Fig.~\ref{fig:prolate}.
%%%%%%%%%%%%%%%%%%%%%%%%%%%%%%%%%%%%%%%%%%%%%%%%%%%%%%%%%%%%%%%%%%%%%%%%%
%             FIG. 8:Intrinsic quadrupole and octupole moments
%%%%%%%%%%%%%%%%%%%%%%%%%%%%%%%%%%%%%%%%%%%%%%%%%%%%%%%%%%%%%%%%%%%%%%%%%
%%%%%%%%%%%%%%%%%%%%%%%%%%%%%%%%%%%%%%%%%%%%%%%%%%%%%%%%%%%%%%%%%%%%%%%%%%%%%%%%%%%%%%

Previously, we have found for the intrinsic quadrupole moment of the proton and 
$\Delta^+$ in the quark model with two-body exchange currents~\cite{Hen01} 
\be
\label{intquad}
Q_0^p = -r^2_n, \qquad Q_0^{\Delta^+}  =  r^2_n. 
\ee
Thus, the intrinsic quadrupole moment of the proton is given
by the negative of the neutron charge radius and is therefore
positive, whereas the intrinsic quadrupole moment of the $\Delta^+$ 
is negative. This corresponds to a prolate proton and an oblate
$\Delta^+$ shape. The quark model with exchange currents also suggests that the nonsphericity of the 
proton charge density is mainly connected with 
collective quark-antiquark degrees of freedom, the distribution of
which has a prolate shape.  

The concept of an intrinsic quadrupole moment of the nucleon can
be generalized to an intrinsic quadrupole charge distribution and a
corresponding form factor~\cite{buc05,Buc07b}.
To show this, we first decompose the proton and neutron charge form factors
in two terms $G_{C,sym}$ and $G_{C,def}$, 
coming from the spherically symmetric and the intrinsic quadrupole 
part of the physical charge density respectively
\bea
\label{voldefdecomp}
G_C^p(Q^2) \!& = &\!G_{C,sym}^p(Q^2) -\frac{1}{6} \, Q^2 \, G_{C, def}(Q^2), \nonumber \\
G_C^n(Q^2)\! & = &\! G_{C,sym}^n(Q^2) + \frac{1}{6} \, Q^2 \, G_{C, def}(Q^2).
\eea
The factor $Q^2$ in front of $G_{C, def}$ arises for dimensional reasons
and guarantees that the normalization of the charge form factors
is preserved. 

In coordinate space this corresponds to the usual multipole
decomposition of the charge density
\be
\rho({\bf r})=\underbrace{\rho_0(r) Y^0_0({\bf r})}_{{\rm monopole}}
+ \underbrace{\rho_2(r)Y^2_0({\bf r})}_{{\rm quadrupole}} + \ldots,
\ee
where the $\rho_0(r)$ part gives rise to $G_{C, sym}(Q^2)$ and the $\rho_2(r)$ 
part is connected with $G_{C, def}(Q^2)$. In terms of fundamental
photon-quark processes depicted in Fig.~\ref{fig:one_two_three} 
the monopole part $G_{C, sym}$ comes from one-quark currents, whereas 
the intrinsic quadrupole part $G_{C, def}$ is
mainly due to two- and three-quark currents.

From the relation between the measurable $N \to \Delta$ quadrupole and 
elastic neutron charge form factors in Eq.(\ref{C2C0rel}) we find for the intrinsic charge quadrupole form factor $G_{C, def}(Q^2)$ of the nucleon
\bea
\label{intrinsicC2ff}
G_{C, def}(Q^2) & = &  -\sqrt{2} G_{C2}^{N \to \Delta}(Q^2)
= \frac{6}{Q^2} G_{C}^n(Q^2), \nonumber \\
G_{C, def}(0) & = & -r_n^2=Q_0^p.
\eea
The zero momentum limit follows from l' Hospital's rule 
and Eq.(\ref{intquad}).
This shows that $G_{C, def}(Q^2)$ as defined in 
Eq.(\ref{intrinsicC2ff}) is the proper generalization of the
intrinsic quadrupole moment $Q_0^p$ to finite momentum transfers.

To exhibit the effect of the intrinsic quadrupole form factor on the elastic
nucleon form factors we insert Eq.(\ref{intrinsicC2ff}) 
into Eq.(\ref{voldefdecomp}) and obtain
\bea
\label{isoscalar}
G_{C}^p(Q^2) & = & G_{C, sym}^p(Q^2)-G_{C}^n(Q^2) \nonumber \\
&  =  & \underbrace{G_C^{IS}(Q^2)}_{{\rm spherical}} - 
\underbrace{G_C^n(Q^2)}_{{\rm deformed}}, \nonumber \\
G_C^n(Q^2)  & = &\frac{1}{6} \, Q^2 \, G_{C, def}(Q^2),
\eea
where the isoscalar nucleon charge form factor is defined as
\be
G_C^{IS}(Q^2)= G_C^p(Q^2)+G_C^n(Q^2).
\ee  
We propose that $G_{C,sym}^n(Q^2)=0$ so that the neutron charge form factor is
solely given by $G_{C, def}(Q^2)$ as stated in Eq.(\ref{isoscalar}).
Thus, the relation between the $N \to \Delta$ and neutron charge form
factors in Eq.(\ref{C2C0rel}) is seen here to have an important
implication for the nucleon ground state itself. 

There are several observable consequences of Eq.(\ref{intrinsicC2ff}) and 
Eq.(\ref{isoscalar}) as discussed in Ref.~\cite{buc05,Buc07b}. 
At low $Q^2$ the nucleon's prolate deformation is 
reflected in a proton charge radius increase by an amount $-r_n^2$,
with respect to $r_{IS}^2$ and by a novel nucleon 
size parameter $r_{def}^2=r^2_{C2}=(7/10) (r_n^4/r_n^2)$ measuring 
the extension of the intrinsic quadrupole charge density.
At intermediate $Q^2$ it leads to the conclusion that 
the dip structure observed in the proton
charge form factor~\cite{fri03} at around $Q^2 \approx 0.3$ GeV$^2$
is due to a corresponding structure in the neutron charge form factor
at the same $Q^2$. Finally, at high $Q^2$ it leads to  
the observed decrease of the charge over magnetic
form factor ratio~\cite{jon00}.

\subsection{$N \to N^*(1680)$ transition quadrupole moment}
\label{subsec:NN*quadmom}
After projecting the charge-charge commutation relation in Eq.(\ref{current_algebra}) onto the Coulomb quadrupole part by multiplying with 
$\sqrt{16 \pi/5} \, r^2 \, Y^2_0(\hat{{\bf r}})$ and integrating over space we get 
\be 
\label{cural1}
\left[ Q_z^{\alpha}, Q_z^{\beta}\right] = i f_{\alpha \beta \gamma} \,
 \int \, d^3{\bf r} (3 z^2 -r^2)^2 \, \rho^{\gamma}({\bf r}),
\ee
where 
\be
\label{cural2}
Q_z^{\alpha}=\int \, d^3{\bf r} (3 z^2 -r^2) \, \rho^{\alpha}({\bf r}).
\ee 
For the flavor (isospin) index we
take $\alpha=1$ and $\beta=2$, entailing $\gamma=3$ and $f_{123}=1$. 
The subscript indicates the $z$ component of the quadrupole tensor.
For calculational convenience we transform in flavor (isospin) space 
from a cartesian to a spherical basis using the ladder operators
\be
\label{sphericalbasis}
Q^{\pm} = \mp \frac{1}{\sqrt{2}}\, (Q^1 \pm i \, Q^2), \ \ \ \ 
Q^{0}  =   Q^3.
\ee
which leads to 
\be 
\left[ Q_z^{+}, Q_z^{-}\right] = -\int \, d^3{\bf r} \,(3 z^2 -r^2)^2 \rho^0({\bf r}).
\ee

%%%%%%%%%%%%%%%%%%%%%%%%%%%%%%%%%%%%%%%%%%%%%%%%%%%%%%%%%%%%%%%%%%%%%%
\begin{table}[b]
\begin{center}
\caption[C2 and M3 moments]{Empirical helicity amplitudes $A_{1/2}$, 
$A_{3/2}$ and $S_{1/2}$ in units $[10^{-3}$ GeV$^{-1/2}]$~\cite{tia16,Tia11,Pat16} 
for the electromagnetic $N \to N^*(1680)$ transition and the corresponding 
quadrupole moment $Q_{N \to N^*(1680)}$ in units [fm$^2$] determined 
from the transverse helicity amplitudes and Siegert's theorem (5th column) or 
directly from the scalar helicity amplitude $S_{1/2}$ (last column).}
\label{tab:helicity_amp}
\begin{tabular}{|l | c | c | c | c | c|}
\hline
& $A_{1/2}$  & $A_{3/2}$ & $S_{1/2}$ & 
$Q_{N \to N^*}$ & $Q_{N \to N^*}$ \\ 
\hline
%%%%%%%%%%%%%%%%%%%%%%%%%%%%%%%%%%%%%%%%%%%%%%%%%%%%%%%%%%%%%%%%%%%%%%%%%%%%%%%%%%%%%%%%
$p\to p^*$ & $-15 \pm 6$  & $\phantom{-}133 \pm 12$ & $-44$ & 0.203(27) &  0.073 \\
%%%%%%%%%%%%%%%%%%%%%%%%%%%%%%%%%%%%%%%%%%%%%%%%%%%%%%%%%%%%%%%%%%%%%%%%%%%%%%%%%%%%%%%%
$n\to n^*$ & $\phantom{-}29 \pm 10$ & $-33 \pm 9$  & $0$ &  $-0.021(27)$  & $0$ \\
%%%%%%%%%%%%%%%%%%%%%%%%%%%%%%%%%%%%%%%%%%%%%%%%%%%%%%%%%%%%%%%%%%%%%%%%%%%%%%
\hline
\end{tabular}
\end{center}
\end{table}
%%%%%%%%%%%%%%%%%%%%%%%%%%%%%%%%%%%%%%%%%%%%%%%%%%%%%%%%%%%%%%%%%%%%%%
For an evaluation between nucleon ground states with orbital angular momentum 
$L=0$, the right-hand side can be simplified as follows  
\be 
\langle p \vert \left[ Q_z^{+}, Q_z^{-}\right] \vert p \rangle  = -\frac{4}{5}
\langle p \vert \int \, d^3{\bf r} \, r^4 \rho^0({\bf r})  \vert p \rangle.
\ee
Note that the right hand side is nonzero even though the proton does not have a
spectrosopic quadrupole moment.
Inserting on the left-hand side a sum of intermediate $N^*$ resonances 
\be
\sum_{N^*} \vert N^*\rangle \langle N^* \vert =1 
\ee
of which only those contribute that can be reached with an orbital angular momentum $L=2$ 
and an isospin $T=1$ operator such as $Q^{\pm}$ 
we obtain a relation between the fourth moment 
of the ground state charge density and a sum of squared transition quadrupole moments.

If we include only the $N^*(1680)$ as intermediate state, we get 
\bea
& & \langle p \vert Q_z^+\vert n^*(1680)\rangle \, \langle n^*(1680) \vert Q_z^-\vert p \rangle \nonumber \\
& = & -2 Q^2_{IV}(p \to p^*(1680)) =  -\frac{4}{5}\, r_{IV}^4(p),
\eea
where we have converted $Q_z^{\pm}$ back to $Q_z^0$ 
using the Wigner-Eckart theorem in isospin space. 
Here, $Q_{IV}(p \to p^*(1680))$ and  $r_{IV}^4(p)$ denote the isovector parts of the 
proton's transition quadrupole moment and fourth radial moment 
This leads to the result
\be
\label{curalg_transquad}
Q_{IV}(p \to p^*(1680))= \sqrt{\frac{1}{5} (r_p^4 -r_n^4)},
\ee
where $r_p^4$ ($r_n^4$) is the fourth radial moment of the proton (neutron) charge distribution. This agrees with Bietti~\cite{Bie66} except for a factor 2 
on the right-hand side not contained in Ref.~\cite{Bie66}.

For numerical evaluation we need the nucleon structure parameters 
$r_p^4$ and $r_n^4$, which are connected with the
curvature of the corresponding charge monopole form factors. 
These are not very well known experimentally.
For example, for the fourth radial moment the following values can be found 
in the literature:
$r_p^4=0.82\pm 1.02$ fm$^4$~\cite{Gri16},
$r_p^4=1.32\pm 0.96$ fm$^4$~\cite{Hig16}, 
$r_p^4=2.01\pm 0.05$ fm$^4$~\cite{Sic17},
$r_p^4=2.59\pm 0.19$ fm$^4$~\cite{Dis11}.

We have calculated the lowest radial moments of the proton and neutron 
using the form factor decomposition in Eq.(\ref{voldefdecomp}), 
where we have used the Galster~\cite{Gal71} parametrization for the neutron charge form factor,
and a dipole form for the isoscalar nucleon form factor
\bea
G_{C}^{n}(Q^2) & = & -  \frac{a \tau}{1+d\tau} \, 
\frac{\mu_n}{(1+Q^2/\Lambda^2_M)^2} \nonumber \\
G_{C}^{IS}(Q^2)& = &(1+Q^2/\Lambda^2_{IS})^{-2}.
\eea
This leads to the radial moments listed in Table~\ref{tab:higherradmom}.
%%%%%%%%%%%%%%%%%%%%%%%%%%%%%%%%%%%%%%%%%%%%%%%%%%%%%%%%%%%%%%%%%%%%%%
\begin{table}[b]
\begin{center}
\caption[higherradialmoments]{Radial moments of the proton and neutron charge distribution based on the form factor decomposition of Eq.(\ref{voldefdecomp})
using the following parameters: $\Lambda^2_M=19.87$ fm$^{-2}$ determined from the experimental proton magnetic radius $r_M=0.777$ fm$^2$~\cite{Ber10}
$\Lambda^2_{IS}=18.43$ fm$^{-2}$ 
determined from the experimental isoscalar radius $r^2_{IS}=0.651$ fm$^2$
and $a=0.9$, $d=1.75$ from a fit to the neutron form factor data~\cite{Gra01}.}
\label{tab:higherradmom}
\begin{tabular}{|l | r | r | r |}
\hline
& $r^2$ [fm$^2$] & $r^4$ [fm$^4$] & $r^6$ [fm$^6$]  \\ 
\hline
%%%%%%%%%%%%%%%%%%%%%%%%%%%%%%%%%%%%%%%%%%%%%%%%%%%%%%%%%%%%%%%%%%%%%%%%
$p$   & $0.7658$     & $1.3356$   &  $4.4070$  \\
%%%%%%%%%%%%%%%%%%%%%%%%%%%%%%%%%%%%%%%%%%%%%%%%%%%%%%%%%%%%%%%%%%%%%%%%
$n$   & $-0.1149$    & $-0.2757$  &  $-1.1866$    \\
%%%%%%%%%%%%%%%%%%%%%%%%%%%%%%%%%%%%%%%%%%%%%%%%%%%%%%%%%%%%%%%%%%%%%%%%
\hline
\end{tabular}
\end{center}
\end{table}
%%%%%%%%%%%%%%%%%%%%%%%%%%%%%%%%%%%%%%%%%%%%%%%%%%%%%%%%%%%%%%%%%%%%%%

Using the empirical helicity amplitudes from PDG~\cite{Pat16} 
and MAID~\cite{tia16,Tia11} listed in Table~\ref{tab:helicity_amp}
and the conversion formulae given in Appendix B 
we obtain for the experimental $N\to N^*(1680)$ transition quadrupole moments
\bea
Q_{p \to p^*(1680)} & = &  0.203(27) \, {\rm fm}^2 \nonumber \\
Q_{n \to n^*(1680)} & = & -0.021(27) \,  {\rm fm}^2
\eea 
based on the transverse helicity amplitudes and Siegert's theorem. 
This gives $Q_{IV}(p \to p^*(1680))= 0.112(27)$ fm$^2$(exptl).
Alternatively, from the scalar helicity amplitudes 
we get $Q_{IV}(p \to p^*(1680))= 0.037$ fm$^2$(exptl).
This has to be compared to 
$Q_{IV}(p \to p^*(1680))= 0.57$ fm$^2$(theory) from Eq.(\ref{curalg_transquad}), 
where we have used $r_p^4=1.34$ fm$^4$ 
and neutron $r_n^4=-0.28$ fm$^4$ from Table~\ref{tab:higherradmom}.

Evidently, the agreement between a theory based on a single intermediate resonance and experiment is fairly bad. But the nucleon spectrum has several more 
resonances~\cite{Cre13,Bur15} with quantum numbers that can be reached by 
an $L=2$ operator, e.g. 
the positive parity states $p^*(1860)$ and $p^*(2000)$ with $J=5/2$ in addition to 
the $p^*(1720)$ and $p^*(1900)$ resonance with $J=3/2$. Including these 
excited states on the commutator side, we obtain  
$Q_{IV}(p \to p^*) \approx 0.25$ fm$^2$(theory). If additional excited states exist, 
the agreement between theory and experiment may further improve. 
Aside from the uncertainty related to the number of excited states,
there is considerable uncertainty with respect to the fourth radial moments 
of the ground state
charge distribution, which are not well known experimentally.

Clearly, more detailed theories achieving better agreement with experiment may be formulated.
In this work, our main point is to study the relationship between 
transition multipole moments and ground state properties. 
The present symmetry based current algebra 
approach makes this connection to some extent transparent. 
In particular, our results suggest 
that the radial moments $r_p^4$ and $r_n^4$ are connected 
with the quadrupole excitation of the $p^*(1680)$ resonance.

\subsection{Magnetic octupole moments of ground state baryons} 
\label{subsec:octmom}
For the construction of a rank $J=3$ magnetic octupole moment operator from the
generators of the group SU(6) group 
%to be used in the GP method of sect.~\ref{sec:gp}
we need a tensor of rank 3 in spin space. This operator
must involve the Pauli spin matrices of three different quarks.
If two of these had the same particle index, the SU(2) spin commutation relations 
would reduce their action to a single Pauli matrix, and  
we could only build a spin tensor of rank 2.
Therefore, a tensor of rank 3 in spin space must necessarily be a 
three-quark operator.

We have previously shown that the magnetic octupole moment operator can 
be constructed from a two-body quadrupole moment operator multiplied 
by the spin of the third quark~\cite{Buc08}
\begin{eqnarray} 
\label{para3}
{\Omega}_{[3]} & = & 
C\sum_{i \ne j \ne k}^3 e_i \left ( 3 \sigma_{i \, z} \sigma_{ j \, z}
-\xbf{ \sigma}_i \cdot \xbf{ \sigma}_j \right )\xbf{ \sigma}_{k}.
\end{eqnarray}

As a three-body operator $\Omega$
transforms according to the {\bf 2695} irrep of SU(6), 
which occurs only once on the right-hand side of Eq.(\ref{baryon56}).
In addition, Eq.(\ref{2695decomp}) shows that the 
flavor {\bf 8}, spin 3 tensor $({\bf 8},{\bf 7})$ 
appears only once in this decomposition.
Hence, for the ${\bf 56}$ dimensional irrep of ground state baryons
there is a unique three-quark magnetic octupole operator.
%%%%%%%%%%%%%%%%%%%%%%%%%%%%%%%%%%%%%%%%%%%%%%%%%%%%%%%%%%%%%%%%%%%%   

The magnetic octupole moments $\Omega_{B^*}$ 
are obtained by calculating the
matrix elements of the octupole operator $\Omega_{[3]}$ between the 
three-quark spin-flavor wave functions $\vert W_{B^{*}} \rangle $
\begin{eqnarray}
\label{mom_matrixelements} 
\Omega_{B^*} & = &\left \langle W_{B^*} \vert  {\Omega}_{[3]} 
\vert W_{B^*} \right \rangle ,
\end{eqnarray}  
where $B^*$ denotes a member of the spin 3/2 baryon decuplet.
For example, sandwiching Eq.(\ref{para3})
between the spin-isospin wave functions of the $\Delta(1232)$ gives
\bea
\label{twothree}
\Omega_{\Delta} & = &\langle W_{\Delta} 
\vert { {\Omega}}_{[3]} 
\vert  W_{\Delta} \rangle  =  4 C e_{\Delta},
\eea
where $e_{\Delta}$ is the $\Delta$ charge.
Similarly, the magnetic octupole moments for the other decuplet baryons
are calculated. In this way Morpurgo's method yields an efficient parameterization
of baryon octupole moments in terms of just one unknown parameter~\cite{Buc08}.

To obtain an estimate for $\Omega_{\Delta^+}$ 
we use the pion cloud model~\cite{Hen01}
where the $\Delta^+$ wave function for maximal spin projection 
is written as 
\be
\vert \Delta^+ J_z=\frac{3}{2}\rangle = 
\beta'\Bigl ( 
\sqrt{\frac{1}{3}} 
\vert n' \pi^+ \rangle
+ 
\sqrt{\frac{2}{3}} 
\vert p' \pi^0 \rangle \Bigr )
\vert \uparrow  Y^1_1(\hat{{\bf r}}_{\pi})  \rangle.
\ee
In this model the magnetic octupole moment operator is a product 
of a quadrupole operator in pion variables and a magnetic
moment operator in nucleon variables
\be 
\Omega_{\pi N} = \sqrt{\frac{16 \pi}{5}} 
\, r_{\pi}^2 \, Y^2_0(\hat{{\bf r}}_{\pi}) \, \, \mu_N \,\tau_z^N \, 
\sigma_z^N,
\ee
where $\mu_N=1/(2M_N)$ is the nuclear magneton.
Here, the spin-isospin structure of $\Omega_{\pi N}$
is infered from the $\gamma \pi N$ and $\gamma \pi$ currents 
of the static pion-nucleon model~\cite{Hen62}.

With these expressions the $\Delta^+$ magnetic octupole moment is readily 
calculated~\cite{Hen01} 
\be
\label{deltamom_pion}
\Omega_{\Delta^+} = -\frac{2}{15} \, {\beta'}^{2}\, r_{\pi}^2 \
\mu_N = Q_{\Delta^+}\, \mu_{N} = r_n^2 \, \mu_N,
\ee
where $Q_{\Delta^+}$ is the $\Delta^+$ quadrupole moment 
and $r_n^2$ the neutron charge radius. With the experimental value 
of the latter and $\mu_N$ expressed in $[{\rm fm}]$  one gets
$\Omega_{\Delta^+} =-0.012\,\,{\rm fm^3}$.
The negative value of $\Omega_{\Delta^+}$ implies that the magnetic moment 
distribution in the $\Delta^+$ is oblate and hence  
has the same geometric shape as the charge distribution as shown in Fig.~\ref{fig:prolate}.

In Table~\ref{tab:octumom} we show our results for the decuplet octupole moments 
of the $\Delta$ and the $\Omega^-$ baryon expressed in terms of the GP constant $C$. 
Results for other decuplet baryons can be found in Ref.~\cite{Buc08}.
%%%%%%%%%%%%%%%%%%%%%%%%%%%%%%%%%%%%%%%%%%%%%%%%%%%%%%%%%%%%%%%%%%%%%%%%%%%%%%%%
\begin{table}[htb]
\begin{center}
\caption[M3 moments]{\label{tab:octumom} 
Magnetic octupole moments of decuplet baryons with 
$C=-0.003$ fm$^3$ as determined from Eq.(\ref{deltamom_pion}) and Eq.(\ref{twothree}). 
SU(3) flavor symmetry breaking is 
characterized by the ratio $r$ of u-quark and s-quark masses 
$r=m_u/m_s=0.6$. From Ref.~\cite{Buc08}.} 
\begin{tabular}{ | l | r |  r |} 
\hline
&  $\Omega$  & $\Omega$\, [fm$^3$]    \\
\hline
$\Delta^{-}$     & $ -4C $	  & $0.012 $           \\
$\Delta^{0}$     &    $ 0 $    	  & 0                   \\
$\Delta^{+}$     & $4C $ 	  & $-0.012 $            \\
$\Delta^{++}$    & $8C $  	  & $-0.024$            \\
$\Omega^-$ & $-4C\,r^3$	  & $0.003 $          \\ 
\hline 
\end{tabular} 
\end{center}
\end{table}

\subsection{Intrinsic magnetic octupole form factor of the nucleon}
\label{subsec:intrinsimom}

It is now fairly certain that the nucleon ground state charge distribution
is not spherically symmetric. The geometric shape of the 
nucleon charge distribution is described by its intrinsic quadrupole 
moment $Q_0$ as discussed in sect.~\ref{subsec:intrinsicqm}. 
It is conceivable that also the spatial current 
distribution and specifically the magnetic moment distribution inside the nucleon 
deviates from spherical symmetry. The existence of a
nonvanishing $\Delta$ magnetic octupole and a fairly large $p \to p^*(1680)$ magnetic 
octupole moment provide some evidence that the nucleon has 
an intrinsic magnetic octupole moment.

Note that the definition for the octupole moment in Eq.(\ref{C2andM3}) is analogous to the one for the charge quadrupole moment if the magnetic moment density $({\bf r} \times {\bf J}({\bf r}))_z$ is replaced
by the charge density $\rho({\bf r})$. 
Thus, the magnetic octupole moment measures 
the deviation of the spatial magnetic moment distribution from
spherical symmetry. Specifically, for a prolate (cigar-shaped) 
magnetic moment distribution $\Omega >0$, 
while for an oblate (pancake-shaped) magnetic moment distribution 
$\Omega <0$ as depicted in Fig.~\ref{fig:prolate}. 
We also see from Eq.(\ref{C2andM3}) that the typical size of a 
magnetic octupole moment is 
\be
\Omega \simeq \mu \, r^2
\ee
where $\mu$ is the magnetic moment and $r^2$ a size parameter 
related to the quadrupole moment of the system.

From Eq.(\ref{deltamom_pion}) and the discussion 
in sect.~\ref{subsec:intrinsicqm} we infer 
\be
\label{intrinsic_oct_mom}
\Omega_0^{\Delta^+}= r_n^2 \, \mu_N,  \qquad \Omega_0^{p}= -r_n^2 \, \mu_N.
\ee
Eq.(\ref{deltamom_pion}) is seen to be the zero-momentum transfer 
limit of the magnetic octupole form factor of the $\Delta^+$ 
\be
G_{M3}^{\Delta^+}(Q^2) = -\frac{6}{Q^2}\, G_{C}^n(Q^2) \, \mu_N. 
\ee

Analogous to the discussion in sect.~\ref{subsec:intrinsicqm} we
 decompose the magnetic dipole form factor of the proton 
in two terms $G_{M, sym}$ and $G_{M, def}$, 
coming from the spherically symmetric and the intrinsic octupole 
part of the magnetic moment density respectively
\be
\label{voldefdecomp_oct}
G_M^p(Q^2) = G_{M, sym}^p(Q^2) -\frac{1}{6} \, Q^2 \, G_{M, def}(Q^2).
\ee
The factor $Q^2$ in front of $G_{M, def}$ arises for dimensional reasons
and guarantees that the magnetic moment of the proton remains unchanged. 
For $G_{M, sym}^p(Q^2)$ we take a dipole form factor 
$G_{M, sym}^p(Q^2)=\mu_p [1+Q^2/\Lambda^2_M]^{-2}$.
For the intrinsic magnetic octupole form factor $G_{M, def}(Q^2)$ we find 
\be
\label{intrinsicM3ff}
G_{M,def}(Q^2)= - G_{M3}^{\Delta^+}(Q^2)= \frac{6}{Q^2} \, G_{C}^n(Q^2) \, \mu_N 
\ee
which gives $G_{M,def}(0)=\Omega_0^p=-r_n^2 \mu_N$ and thus 
shows that $G_{M,def}(Q^2)$ is the proper generalization of 
Eq.(\ref{intrinsic_oct_mom}) to finite momentum transfers.

To exhibit the effect of the intrinsic octupole form factor on the
magnetic dipole form factor of the nucleon we insert Eq.(\ref{intrinsicM3ff}) 
into Eq.(\ref{voldefdecomp_oct}) and obtain
\bea
\label{M1M3decomp}
G_{M}^p(Q^2) & = & G_{M,sym}^p(Q^2)-G_{C}^n(Q^2) \mu_N.
\eea  
At low $Q^2$ the proton's prolate magnetic dipole distribution 
leads to a small magnetic radius increase by an amount $-r_n^2 \, \mu_N/\mu_p$ 
relative to the symmetric part given by $12/\Lambda^2_M$.
At intermediate $Q^2$ our finding suggests that 
the dip structure observed in the proton
magnetic form factor~\cite{fri03} at around $Q^2 \approx 0.3$ GeV$^2$
is due to a corresponding structure in the neutron charge form factor
at the same $Q^2$. 

%===========================================================================
\subsection{$N \to N^*(1680)$ transition octupole moment}
\label{subsec:NN*octmom}
After projecting the current-current commutation relation in Eq.(\ref{current_algebra}) onto the magnetic octupole parts of the currents  
according to Eq.(\ref{C2andM3}) we obtain
\bea
\label{comrel1} 
\left [\tilde{\Omega}^1_z, \tilde{\Omega}^2_z\right ]
 & = & i\,\left (\frac{3}{8} \right )^2 
(2 \, M_N)^2   \nonumber \\
 & & \int d^3 r \, (3 z^2- r^2)^2 \,  (x^2+y^2)\,  \rho^3({\bf r}), 
\eea
where $\rho^3({\bf r})$ is the isovector component of the charge density.
The axial current term on the right hand side in Eq.(\ref{current_algebra}) does
not contribute here because by definition 
the magnetic octupole moment operators are evaluated for the $z$ component
that is for $i=j=3$.
%%%%%%%%%%%%%%%%%%%%%%%%%%%%%%%%%%%%%%%%%%%%%%%%%%%%%%%%%%%%%%%%%%%%%%

Converting to a spherical basis in isospin space analogous to
 Eq.(\ref{sphericalbasis}) and  
sandwiching Eq.(\ref{comrel1}) between proton ground states we obtain
\bea
\label{comrel2} 
& & \langle p \vert \tilde{\Omega}^+_z \vert n^*(1680) \rangle 
\langle n^*(1680) \vert \tilde{\Omega}^-_z \vert p \rangle \nonumber \\
  & = & \frac{3}{56} \,  \, 
(2 \, M_N)^2 \, \, \langle p \vert 
\int d^3 r  \, \, r^6  \, \rho^0({\bf r}) \, \vert p \rangle 
\eea
where we have included only the $n^*(1680)$ intermediate state with spin $5/2$
and isospin $1/2$.
With the help of the Wigner-Eckart theorem  the left-hand side can
be expressed in terms of the isovector part of the $p \to p^*(1680)$ 
transition octupole moment. 
Eq.(\ref{comrel2}) then provides a relation between the isovector 
$p \to p^*(1680)$ transition octupole moment
and the sixth moments of the proton and neutron charge distributions
\be
\label{comrel3}
\Omega_{IV}(p \to p^*(1680))  
=\sqrt{\frac{3}{224} \left( r^6_{p}-r^6_n \right)}. 
\ee
A numerical estimate for $\Omega_{IV}(p \to p^*(1680))$ 
based on the radial moments in Table~\ref{tab:higherradmom} gives 
$\Omega_{IV}(p \to p^*(1680))=0.27$ fm$^3$(theory) for a single resonance.
If the 2 additional excited $J=5/2$ states seen in the spectrum 
are included on the commutator side we get 
$\Omega_{IV}(p \to p^*) \approx 0.16$ fm$^3$(theory).

To compare our theory with experiment,
we use the conversion formulae in Appendix B and 
the helicity amplitudes in Table~\ref{tab:helicity_amp} and get
\bea
\Omega_{p \to p^*(1680)} & = &  0.164(22) \, {\rm fm}^3, \nonumber \\ 
\Omega_{n \to n^*(1680)} & = & -0.079(25) \, {\rm fm}^3.
\eea
This gives for the isovector term of the proton transition octupole moment 
$\Omega_{IV}(p \to p^*(1680))=0.121(23)$ fm$^3$(exptl) compared to 
$\Omega_{IV}(p \to p^*(1680))=0.27$ fm$^3$(theory).

Finally, we calculate the $p \to p^*(1680)$ transition multipole ratio
\be
\frac{C2}{M3}(Q^2=0):=  \frac{15}{4 \vert {\bf q} \vert} \,
\frac{Q_{p \to p*(1680)}}{\Omega_{p \to p^*(1680)}},
\ee
which gives for the isovector part of the $N \to N^*$ transition
$(C2/M3)_{IV}=2.7$(theory) compared to  $(C2/M3)_{IV}=1.2$(exptl).

\section{Summary} 

We have used several SU(6) symmetry based methods to calculate the charge quadrupole and magnetic octupole moments of selected members of the baryon 56 dimensional spin-flavor supermultiplets with orbital angular momentum $L=0$ and $L=2$ and compared our results to experiment.

We have shown that quadrupole and octupole moments receive only contributions from 
second and third order symmetry breaking connected with two-quark and 
three-quark currents. This provides a unique opportunity to get 
information on the sign and magnitude of these two- and three-quark exchange currents,
which describe $q\overline{q}$ and gluon degrees of freedom in the nucleon.

More importantly, the symmetry based methods used here reveal that there are 
interesting relations between the transition multipole moments 
and the radial moments of the ground state charge distribution.
In the light of the present investigation, the sign and size of the 
radial moments contain important information on the angular shape of the 
nucleon ground state. 

Finally, we have extracted the intrinsic charge quadrupole and for the first
time the intrinsic magnetic octupole form factors of the nucleon from
empirical $N \to \Delta$ transition form factors. Our results show that 
these intrinsic form factors produce observable deviations
from a smooth dipole behavior in the proton elastic form factors.

\centerline{{\bf Appendix A}}

For Eq.(\ref{qmnr}) to be valid we have to show that $B'=-B/2$ and $C'=-C/2$.
Here we show using only group theoretical arguments that  $B'=-B/2$.
At the end of sect.~\ref{sec:spin-flavor} we have mentioned that 
the spin scalar $({\bf 8,1})$ 
and spin tensor $({\bf 8,5})$ operators belong to the same SU(6) irreps
and that their matrix elements are related by an SU(6) Clebsch-Gordan coefficient.
Using the notation of sect.~\ref{sec:spin-flavor} we write for the charge radius 
and quadrupole operators
\bea
r^2 & = & \phantom{-\sqrt{5}} \, \, 
\Omega^{\bf 405}_{({\bf 8}, {\bf 1})} + \Omega^{\bf 2695}_{({\bf 8}, {\bf 1})}
\nonumber \\
\mathcal{Q}  & = &  -\sqrt{5} \, \,\left(
 \Omega^{\bf 405}_{({\bf 8}, {\bf 5})} + \Omega^{\bf 2695}_{({\bf 8}, {\bf 5})}\right).
\eea 
Both operators are recognized here as different components of common 
SU(6) tensor operators $\Omega^{\bf 405}$ and $\Omega^{\bf 2695}$. 

According to the generalized Wigner-Eckart theorem, 
the matrix elements of $\Omega^{\bf 405}$ and $\Omega^{\bf 2695}$ evaluated between 
the ${\bf 56}$ multiplet can be factorized into 
a common reduced matrix element (indicated by a double bar), 
which is the same for the entire 
multiplet, and an SU(6) Clebsch-Gordan (CG) coefficient 
\be 
\label{WET}
\mathcal{M}   =  
\langle {\bf 56}_{\nu_f} \vert \, 
\Omega^{{\bf R}}_{\nu} \, \vert {\bf 56}_{\nu_i} \rangle
=
\langle {\bf 56} \vert \vert \, 
\Omega^{{\bf R}} \, \vert \vert {\bf 56} \rangle \, \, 
\left (\begin{matrix}
{\bf 56} &  {\bf R} & {\bf 56} \\
\nu_i & \nu & \nu_f
\end{matrix} \right ), 
\ee
where ${\bf R}$ stands for the ${\bf 405}$ and ${\bf 2695}$ irreps. 

The SU(6) CG coefficients provide relations between the matrix elements of 
different components of the irreducible tensor operator 
$\Omega^{\bf R}_{\nu}$ and the individual states of the
${\bf 56}$ dimensional baryon ground state supermultiplet, 
which are labelled by $\nu_{i}$ and $\nu_{f}$. 
Because SU(6) is a rank five group, the label $\nu$ comprises 
five quantum numbers to uniquely specify a state, three
for SU(3), e.g. total isospin $T$, isospin projection $T_z$,
and hypercharge $Y$, and two for SU(2), e.g. total angular momentum $J$
and its projection $J_z$.

The SU(6) CG coefficient can be split into a unitary scalar 
factor $f_{(\mu,s)}^{{\bf R}}$ 
and a product of SU(3)$_F$ and SU(2)$_J$ CG coefficients as
\be 
\left (\begin{matrix}
{\bf 56} &  {\bf R} & {\bf 56} \\
\nu_f & \nu & \nu_i
\end{matrix} \right )
 =  f^{{\bf R}}_{(\mu, \, s)}  
\left (\begin{matrix}
\xbf{\mu}_f &  \xbf{\mu} & \xbf{\mu}_i \\
\rho_f & \rho & \rho_i 
\end{matrix} \right ) 
 \left(J_i\, J_{i,z} J \, J_{z} \vert J_f \, J_{f,z} \right ),
\ee
where $\mu$ and $s=2J+1$ denote the dimensionalities of the
SU(3) and SU(2) reps. The SU(3)$_F$ CG coefficient label 
$\rho$ comprises the three quantum numbers 
$\rho=(Y, T, T_z)$. Note that the 
SU(6) scalar factor $f^{{\bf R}}_{(\mu, s)}$, 
depends only on the dimensionalities
of the SU(6), SU(3)$_F$ and SU(2)$_J$ irreps involved but not on the SU(3) 
and SU(2) labels $\rho$ and $J_z$.

To prove $B'=-B/2$, consider the two SU(6) matrix elements, which are of interest here
\bea 
\label{grouptheory1}
r_n^2   & = & \langle {\bf 56}_{n} \vert \, 
\Omega^{{\bf 405}}_{({\bf 8},\,{\bf 1}) } \, \vert {\bf 56}_{n} 
\rangle  \nonumber \\
 & =  &  r \ \left ( -\frac{2}{\sqrt{10}} \right ) \, 
\left \lbrack \frac{1}{\sqrt{3}}\,\left (-\sqrt{\frac{1}{20}} \right ) 
-\sqrt{\frac{3}{20}}  \right \rbrack 
=  r \,  \frac{2\sqrt{6}}{15} \nonumber \\
\eea
\bea
\label{grouptheory2}
Q_{p \to \Delta^+}  \! & = & \! -\sqrt{5}  
\langle {\bf 56}_{\Delta^+} \vert \, 
\Omega^{{\bf 405}}_{({\bf 8},\,{\bf 5}) } 
\, \vert {\bf 56}_{p} \rangle  \nonumber \\
& \!= \! & -\sqrt{5}\, r \left ( \frac{1}{\sqrt{10}} \right )
\left \lbrack \frac{2}{\sqrt{15}} \right \rbrack 
\left (- \frac{2}{\sqrt{10}} \right ) 
 =  r  \frac{2\sqrt{3}}{15}, \nonumber \\
\eea
where $r=\langle {\bf 56} \vert \vert \, 
\Omega^{{\bf 405}} \, \vert \vert {\bf 56} \rangle $
is the SU(6) reduced matrix element. 
The SU(3)$_F$ flavor~\cite{mcn65} and SU(2)$_J$ spin CG coefficients
are explicitly shown. In the case of the neutron charge radius, 
the two terms in the brackets refer to SU(3) CG with 
sublabels $\rho=(0,0,0)$ and $\rho=(0,1,0)$ corresponding to the isosinglet 
and isotriplet piece in Eq.(\ref{GMNR}).  As usual, the 
isosinglet part is multiplied by $1/\sqrt{3}$. 
The factor of -2 
between the rank 0 (charge monopole) and rank 2 (charge quadrupole) 
tensors is reflected by the SU(6) scalar factors~\cite{Leb95,coo65} 
$f^{{\bf 405}}_{(8,1)}=-2/\sqrt{10}$ and $f^{{\bf 405}}_{(8,5)}=1/\sqrt{10}$. 
From Eq.(\ref{grouptheory1}) and Eq.(\ref{grouptheory2}) we obtain Eq.(\ref{qmnr}).

For $\Omega^{{\bf 2695}}$ a similar analysis may be done.
Because there are two  $\Omega^{{\bf 2695}}_{(8,5)}$ operators as reflected 
by the multiplicity of the ${\bf (8,5)}$ component in Eq.(\ref{2695decomp}), orthogonal
linear combinations of them must be formed to construct the proper 
quadrupole tensor $\Omega^{{\bf 2695}}_{(8,5)}$ appearing in Eq.(\ref{grouptheory2})~\cite{coo65}. 

\centerline{\bf Appendix B}

For the conversion of the $N \to N^*(1680)$ helicity amplitudes $A_{1/2}$ and $A_{3/2}$
into transition multipole moments defined as in Eq.(\ref{C2andM3}) we have used the following relations
\bea
Q_{p \to p^*(1680)} &\!\!=\!\! &\sqrt{\frac{\omega}{\pi}}\, \frac{2}{e}\, 
\frac{1}{\vert {\bf q}^2 \vert} \, (A_{1/2} + \sqrt{2} A_{3/2}) \nonumber \\
& &  \\
\Omega_{p \to p^*(1680)} &\!\!= \!\! &
\sqrt{\frac{\omega}{2 \pi}}\, \frac{1}{e}\, 
\frac{30 M_N}{\sqrt{2}\vert {\bf q}^3 \vert} \, (A_{1/2} - \frac{1}{\sqrt{2}} A_{3/2}),  \nonumber 
\eea
where we have employed Siegert's theorem to convert the transverse electric quadrupole
moment into a charge quadrupole moment. 
Alternatively, we may convert the scalar helicity 
amplitude $S_{1/2}$ into $Q_{p \to p^*}$ directly to obtain the charge 
quadrupole transition moment
\be
Q_{p \to p^*(1680)} \!\!=\!\!
-\frac{2}{3} \frac{1}{e} \sqrt{\frac{\omega}{2\pi}} \frac{6}{{\bf q}^2}  S_{1/2},
\ee
with $e=1/\sqrt{137}$, 
$\omega=\vert {\bf q}\vert=(M^2_{N^*}-M^2_N)/(2M_{N^*})=0.578$ GeV,
and $M_N=0.939$ GeV.

%%%%%%%%%%%%%%%%%%%%%%%%%%%%%%%%%%%%%%%%%%%%%%%%%%%%%%%%%%%%%%%%%%%%%%%%%

\end{document}